\documentclass[aps]{revtex4}
\usepackage{amsfonts}
\usepackage{amsmath}
\usepackage{amssymb}
\usepackage{graphicx}

\input{tcilatex}

\begin{document}

\title[Bound and resonance states in impure graphene]{Bound and resonance
electron states in the monolayer graphene with the short-range impurities}
\author{Natalie E. Firsova}
\affiliation{Institute for Problems of Mechanical Engineering, the Russian Academy of
Sciences, St. Petersburg 199178, Russia}
\author{Sergey A. Ktitorov}
\affiliation{A.F. Ioffe Physical-Technical Institute, the Russian Academy of Sciences,
St. Petersburg, Russia}
\author{Philip A. Pogorelov}
\affiliation{St. Petersburg State University, Pervogo Maya str. 100, Petrodvoretz, St.
Petersburg 198504, Russia}
\keywords{Dirac equation, delta function, resonance}

\begin{abstract}
Bound and resonance electronic states in impure graphene are studied.
Short-range perturbations for defects and impurities of the types "local
chemical potential" and "local gap" are taken into account. Zero gap and
non-zero gap kinds of graphene are considered. A qualitative analysis of the
electronic spectrum general features and numeric calculations are presented.
A dependence of the resonance widths on the gap value, angular momentum
values and the perturbation amplitudes is investigated.
\end{abstract}

\pacs{81.05.Uw 72.10-d 73.63.-b 73.40.-c}
\maketitle

%\preprint{HEP/123-qed}

%\volumeyear{year}
%\volumenumber{number}
%\issuenumber{number}
%\eid{identifier}
%\date[Date text]{date}
%\received[Received text]{date}

%\revised[Revised text]{date}

%\accepted[Accepted text]{date}

%\published[Published text]{date}

%\startpage{101}
%\endpage{102}
%\tableofcontents

\section{Introduction}

The Dirac equation is one of keystones of the relativistic field theory.
However, it is an important model in the non-relativistic solid state theory
as well. Superconductors with $d-$pairing \cite{d}, the Cohen-Blount
two-band model of narrow-gap semiconductors \cite{keldysh}, \cite{tamar},
electronic spectrum of the carbon tubes form an incomplete list of the
non-relativistic applications of this equation. During the last two years
extremely much attention was payed to the problem of the electronic spectrum
of graphene (see a review \cite{novosel}). Two-dimensional structure of it
and a presence of the cone points in the electronic spectrum make actual a
comprehensive study of the external fields effect on the spectrum and other
characteristics of the electronic states described by the Dirac equation in
the 2+1 space-time. We consider in this work the resonance and bound states
of the 2+1 Dirac equation due to the short-range perturbation using some
approaches developed in Ref.\cite{firs} .We do not take into account the
inter-valley transitions. Particular attention to this case stems from the
effectiveness of short-range scatterers in contrast to the long-range ones:
an effect of the latter is suppressed by the Klein paradox \cite{beenakker}.
Our work takes into account the obvious fact that the Kohn-Luttinger matrix
elements of the short-range perturbation calculated on the upper and lower
band wave functions are not equal in a general case. This means that not
only the potential but the mass perturbation can be present in the perturbed
Dirac equation. Resonance states in the zero-mass Dirac model of graphene
were considered for distinct perturbations in \cite{basko}, \cite{novikov}, 
\cite{matulis} .

The paper is organized as follows. In the first section we formulate the
problem in terms of the 2+1 Dirac equation with the delta function
perturbation taking into account a local change of the potential and mass
(gap) due to the crystal defect. An exact solution of this problem leads us
to the characteristic equation determining the bound and resonance states in
graphene. In the second section we analyze the characteristic equation for
the zero-gap graphene case. Both local and global behavior of resonance
states for various magnitudes of the perturbations are investigated.
Analytic and\ numeric results are presented. Just analytic approach allowed
us to investigate the case of exponentially narrow resonance that could be
difficult to do numerically. In the third section we consider the bound and
resonance states in graphene with non-zero mass (gap). The exponential
approach of the bound state energy levels to the band edge was studied
analytically; numerical analysis is presented as well.

\section{Characteristic equation}

The Dirac equation describing electronic states in graphene reads \cite%
{novosel} 
\begin{equation}
\left( -i\hbar v_{F}\sum_{\mu =1}^{2}\gamma _{\mu }\partial _{\mu }-\gamma
_{0}\left( m+\delta m\right) v_{F}^{2}{}\right) \psi =\left( E-V\right) \psi
,  \label{diracgeneral}
\end{equation}%
where $v_{F}$ is the Fermi velocity of the band electrons, $\gamma _{\mu }$
are the Dirac matrices%
\begin{equation*}
\gamma _{0}=\sigma _{3},\text{ }\gamma _{1}=\sigma _{1},\text{ }\gamma
_{2}=i\sigma _{2},
\end{equation*}%
$\sigma _{i}$ are the Pauli matrices, $2mv_{F}{}^{2}=E_{g}$ is the
electronic bandgap, $\psi \left( \mathbf{r}\right) $ is the two-component
spinor. The electronic gap can appear in the graphene monatomic film lying
on the substrate because of the sublattices mutual shift \cite{gap}; the
spin-orbit interaction can be the reason too. The spinor structure takes
into account the two-sublattice structure of graphene.$\ \delta m\left( 
\mathbf{r}\right) $ and $V(\mathbf{r})$ are the local perturbations of the
mass (gap) and the chemical potential. A local mass perturbation can be
induced by defects in a graphene film or in the substrate \cite{gap}. We
consider here the delta function model of the perturbation:%
\begin{equation}
\delta m\left( \mathbf{r}\right) =-b\delta (r-r_{0}),\text{ }V(\mathbf{r)}%
=-a\delta (r-r_{0}),  \label{delta}
\end{equation}%
where $r$ and $r_{0}$ are respectively the polar coordinate radius and the
perturbation radius. Such short-range perturbation was used in the
(3+1)-Dirac problem for narrow-gap and zero-gap semiconductors in \cite%
{tamar}.

The perturbation matrix elements 
\begin{equation}
diag(V_{1},V_{2})\delta (r-r_{0})  \label{diag}
\end{equation}%
are related to the $a,$ $b$ parameters as follows 
\begin{equation}
V_{1}=-\left( a+b\right) ,\text{ }V_{2}=b-a  \label{abVrelation}
\end{equation}

The delta function perturbation is the simplest solvable short-range model.
Finite radius $r_{0}$ plays a role of the regulator and is necessary in
order to exclude deep states of the atomic energy scale. The finite
perturbation radius $r_{0}$ leads to the quasi-momentum space form-factor
proportional to the Bessel function that justifies our neglect of
transitions between the points $K$\ and $K^{\prime }$ Ref. \cite{tamar}. The
two-dimensional Dirac problem with the scalar short-range perturbation Eq. (%
\ref{delta}) (but without the mass perturbation) was considered in Ref. \cite%
{dong}. The obtained there characteristic equation for the discrete energy
spectrum contains one mistake. We corrected it in our previous work Ref. 
\cite{we} and took into account the mass perturbation $\delta m\left( 
\mathbf{r}\right) .$ Here we present more detailed analysis of the
electronic bound states and new results on resonance states in the zero-gap
and non-zero-gap graphene. Particular attention is paid to the effect of
relative intensities of two perturbations $V_{1}$ and $V_{2}$ ($a$ and $b$).

Let us present the two-component spinor in the form%
\begin{equation}
\psi _{j}(\mathbf{r},t)=\frac{\exp \left( -iEt\right) }{\sqrt{r}}\left( 
\begin{array}{c}
f_{j}\left( r\right) \exp \left[ i\left( j-1/2\right) \varphi \right] \\ 
\\ 
g_{j}\left( r\right) \exp \left[ i\left( j+1/2\right) \varphi \right]%
\end{array}%
\right) ,  \label{spinor}
\end{equation}%
where $j$ is the pseudospin quantum number; $j=\pm 1/2,$ $\pm 3/2,\ldots $.
In opposite to the relativistic theory, this quantum number has nothing to
do with the real spin and indicates a degeneracy in the biconic Dirac point.
The upper $f_{j}\left( r\right) $ and lower $g_{j}\left( r\right) $
components of the spinor satisfy the equations set 
\begin{equation}
\frac{dg_{j}}{dr}+\frac{j}{r}g_{j}-\left( E-m\right) f_{j}=\left( a+b\right)
\delta (r-r_{0})f_{j},  \label{componenteq1}
\end{equation}

\begin{equation}
-\frac{df_{j}}{dr}+\frac{j}{r}f_{j}-\left( E+m\right) g_{j}=\left(
a-b\right) \delta (r-r_{0})g_{j}.  \label{componeq2}
\end{equation}%
These equations have a symmetry: 
\begin{equation}
f_{j}\leftrightarrow g_{j},\text{ }E\rightarrow -E,\text{ }j\rightarrow -j,%
\text{ }a\rightarrow -a.  \label{symm}
\end{equation}%
Let us introduce the function $\varphi _{j}\left( r\right) \equiv
f_{j}/g_{j}.$ It satisfies the equation:%
\begin{equation}
\left[ \frac{d\varphi _{j}}{dr}-\frac{2j}{r}\varphi _{j}-E\left( \varphi
_{j}^{2}+1\right) \right] /\left[ \left( a+b\right) \varphi _{j}^{2}+\left(
a-b\right) \right] +\delta (r-r_{0})=0  \label{phi}
\end{equation}%
Integrating in the vicinity of $r=r_{0}$%
\begin{equation}
\lim_{\epsilon \rightarrow 0}\int_{\varphi _{j}(r_{0}-\delta )}^{\varphi
_{j}(r_{0}+\delta )}\frac{d\varphi _{j}}{\left( a+b\right) \varphi
_{j}^{2}+\left( a-b\right) }=-1,  \label{match1}
\end{equation}%
we obtain the matching condition \ 

\begin{equation}
\arctan \left( \varphi _{j}^{-}\sqrt{\left( a+b\right) /\left( a-b\right) }%
\right) -\arctan \left( \varphi _{j}^{+}\sqrt{\left( a+b\right) /\left(
a-b\right) }\right) =\sqrt{a^{2}-b^{2}},\text{ \ \ \ }a^{2}>b^{2},
\label{match2}
\end{equation}%
where $\varphi _{j}^{-}\equiv \varphi _{j}\left( r_{0}-\delta \right) ,$ $%
\varphi _{j}^{+}\equiv \varphi _{j}\left( r_{0}+\delta \right) ,$ $\delta
\longrightarrow 0$ $.$ The upper and lower component matching conditions
resulting from Eq. (\ref{match2}) read%
\begin{equation}
\left( 
\begin{array}{c}
f_{j}^{+} \\ 
g_{j}^{+}%
\end{array}%
\right) =\overset{\wedge }{A}\left( 
\begin{array}{c}
f_{j}^{-} \\ 
g_{j}^{-}%
\end{array}%
\right) ,  \label{matrixrelation}
\end{equation}%
where 
\begin{equation}
\overset{\wedge }{A}=\left( 
\begin{array}{cc}
\cos \sqrt{a^{2}-b^{2}}, & -\sqrt{\frac{a+b}{a-b}}\sin \sqrt{a^{2}-b^{2}} \\ 
\sqrt{\frac{a-b}{a+b}}\sin \sqrt{a^{2}-b^{2}}, & \cos \sqrt{a^{2}-b^{2}}%
\end{array}%
\right) ,\text{ }a^{2}-b^{2}>0  \label{matrix}
\end{equation}%
is the orthogonal rotation matrix$.$ It transmutes into the orthogonal boost
matrix for $b^{2}>a^{2}$ 
\begin{equation}
\overset{\wedge }{A=}\left( 
\begin{array}{cc}
\cosh \sqrt{b^{2}-a^{2}}, & -\sqrt{\frac{b+a}{b-a}}\sinh \sqrt{b^{2}-a^{2}}
\\ 
-\sqrt{\frac{b-a}{b+a}}\sinh \sqrt{b^{2}-a^{2}}, & \cosh \sqrt{b^{2}-a^{2}}%
\end{array}%
\right) ,\text{ \ \ }b^{2}-a^{2}>0.  \label{matrix2}
\end{equation}%
The general solution can be found solving the second-order equation obtained
by excluding one of the spinor components from the equation set Eq. (\ref%
{componenteq1}), Eq. (\ref{componeq2}) in the domains $0<r<r_{0}$ and $%
r>r_{0}:$%
\begin{equation}
\frac{d^{2}f_{j}}{dr^{2}}+\left[ E^{2}-m^{2}-\frac{j\left( j-1\right) }{r^{2}%
}\right] f_{j}=0.  \label{secondorder}
\end{equation}%
This equation is related to the Bessel one. We assume at first $E$ to be
real and satisfying the inequality $E^{2}<m^{2}.$ Then the general solution
of Eq. (\ref{secondorder}) reads%
\begin{equation}
f_{j}=C_{1}\sqrt{r}I_{j-1/2}\left( \kappa r\right) +C_{2}\sqrt{r}%
K_{j-1/2}\left( \kappa r\right) ,  \label{general}
\end{equation}%
where $\kappa =\sqrt{m^{2}-E^{2}}$ is the principal value of the root; $%
I_{\nu }\left( z\right) $ and $K_{\nu }\left( z\right) $ are the modified
Bessel functions. The constant $C_{2}=0$ in the domain $0<r<r_{0}$, while $%
C_{1}=0$ in the domain $r>r_{0}$. It is useful to introduce the following
notations: 
\begin{equation}
\mathcal{K}_{j}\left( z\right) =K_{j-1/2}\left( z\right) /K_{j+1/2}\left(
z\right) ,\text{ }\mathcal{I}_{j}\left( z\right) =I_{j-1/2}\left( z\right)
/I_{j+1/2}\left( z\right)  \label{ratio1}
\end{equation}%
Expressing the $g_{j}$-component from Eq. (\ref{componeq2}), we can write%
\begin{equation}
\varphi _{j}^{-}=\sqrt{\left( m+E\right) /\left( m-E\right) }\mathcal{I}%
_{j}\left( \kappa r_{0}\right) ,  \label{phi+}
\end{equation}

\begin{equation}
\varphi _{j}^{+}=\sqrt{\left( m+E\right) /\left( m-E\right) }\mathcal{K}%
_{j}\left( \kappa r_{0}\right) .  \label{phi-}
\end{equation}%
Substituting the expressions Eq.(\ref{phi+}), Eq.(\ref{phi-}) into the
matching condition Eq. (\ref{match2}), we obtain the characteristic equation
for the bound state energy levels (for $E^{2}-m^{2}<0$):

\begin{equation}
\kappa \left[ \mathcal{I}_{j}\left( \kappa r_{0}\right) -\mathcal{K}%
_{j}\left( \kappa r_{0}\right) \right] =T\left( a,b\right) \left[
(m-E)\left( a-b\right) +\left( a+b\right) (m+E)\mathcal{I}_{j}\left( \kappa
r_{0}\right) \mathcal{K}_{j}\left( \kappa r_{0}\right) \right]
\label{character1}
\end{equation}%
where $T\left( a,b\right) .$is determined as follows:

\begin{equation}
T\left( a,b\right) =\left\{ 
\begin{tabular}{l}
$\tan \left( \sqrt{a^{2}-b^{2}}\right) /\sqrt{a^{2}-b^{2}}$ \ \ if $%
a^{2}>b^{2},$ \\ 
\\ 
$\tanh \left( \sqrt{b^{2}-a^{2}}\right) /\sqrt{b^{2}-a^{2}}$ \ \ \ if $%
b^{2}>a^{2}.$%
\end{tabular}%
\right.  \label{T}
\end{equation}%
We presented above an analysis for the case of $a^{2}>b^{2};$ the case of $%
b^{2}>a^{2}$ can be considered similarly.

\ This equation turns to the characteristic one obtained in \cite{dong}, for 
$b=0$ apart from the mistakenly omitted terms in the right hand side of Eq. (%
\ref{character1}). This characteristic equation is unambiguously determined
for the bound states with energy levels lying in the real axis segment $%
\left[ -m,\text{ }m\right] .$

We write this equation in another form making the symmetry Eq. (\ref{symm})
manifest:

\begin{eqnarray}
&&\kappa \left[ I_{j-1/2}\left( \kappa r_{0}\right) K_{j+1/2}\left( \kappa
r_{0}\right) -K_{j-1/2}\left( \kappa r_{0}\right) I_{j+1/2}\left( \kappa
r_{0}\right) \right]  \notag \\
&=&T\left( a,b\right) \left[ (m-E)\left( a-b\right) I_{j+1/2}\left( \kappa
r_{0}\right) K_{j+1/2}\left( \kappa r_{0}\right) +\left( a+b\right)
(m+E)I_{j-1/2}\left( \kappa r_{0}\right) K_{j-1/2}\left( \kappa r_{0}\right) %
\right] ,  \label{character1a}
\end{eqnarray}%
This equation was derived for a study of bound states situated in the gap
that will be done in section IV.

An analytic continuation of this equation from the energy real axis segment $%
\left( -m,\text{ }m\right) $ onto the energy complex plane is necessary in
order to study resonance states. Using the obvious relations 
\begin{equation*}
k\equiv \sqrt{m^{2}-E^{2}=}\left\{ 
\begin{array}{c}
-ip,\text{ \ \ if }E>m \\ 
\\ 
ip,\text{ \ \ if }E<-m,%
\end{array}%
\right.
\end{equation*}

where $p=\sqrt{E^{2}-m^{2}},$and the known relations between the Bessel
functions \cite{stegun}

\begin{eqnarray}
I_{\nu }\left( z\right) &=&\left\{ 
\begin{array}{c}
\exp \left( -i\pi \nu /2\right) J_{\nu }\left( z\exp \left( i\pi /2\right)
\right) ,\text{ \ \ }-\pi <\arg z<\pi /2, \\ 
\\ 
\exp \left( i\pi \nu /2\right) J_{\nu }\left( z\exp \left( -i\pi /2\right)
\right) ,\text{ \ \ }-\pi /2<\arg z<\pi ,%
\end{array}%
\right.  \label{besselrelation} \\
K_{\nu }\left( z\right) &=&\left\{ 
\begin{array}{c}
i\pi /2\exp \left( i\pi \nu /\right) H_{\nu }^{\left( 1\right) }\left( z\exp
\left( i\pi /2\right) \right) ,\text{ \ \ }-\pi <\arg z<\pi /2, \\ 
\\ 
-i\pi /2\exp \left( -i\pi \nu /2\right) H_{\nu }^{\left( 2\right) }\left(
z\exp \left( -i\pi /2\right) \right) ,\text{ \ \ }-\pi /2<\arg z<\pi ,%
\end{array}%
\right.  \label{hankelrelation}
\end{eqnarray}

we obtain the characteristic equation in the form%
\begin{equation}
p\left[ \mathcal{J}_{j}\left( pr_{0}\right) +\mathcal{H}_{j}^{\left( \alpha
\right) }\left( pr_{0}\right) \right] =-T\left( a,b\right) \left[ \left(
E-m\right) \left( a-b\right) -\left( E+m\right) \left( a+b\right) \mathcal{J}%
_{j}\left( pr_{0}\right) \mathcal{H}_{j}^{\left( \alpha \right) }\left(
pr_{0}\right) \right] ,  \label{charbesselHJ}
\end{equation}

where 
\begin{equation*}
\alpha =\left\{ 
\begin{array}{c}
1\text{ \ \ if }\func{Re}E>m, \\ 
\\ 
2\text{ \ \ if }\func{Re}E<-m.%
\end{array}%
\right.
\end{equation*}

We have introduced the notation:

\begin{equation}
\mathcal{J}_{j}\left( z\right) =J_{j-1/2}\left( z\right) /J_{j+1/2}\left(
z\right) ,\text{ \ \ }\mathcal{H}_{j}^{\left( \alpha \right) }\left(
z\right) =H_{j-1/2}^{\left( \alpha \right) }\left( z\right)
/H_{j+1/2}^{\left( \alpha \right) }\left( z\right) .  \label{ratio2}
\end{equation}

The cases of $a^{2}=b^{2}$ (coordinate angles bisectrices in the ($a,$ $b$%
)-plane) are degenerate: the energy imaginary part vanishes in this limit
since according to Eq. (\ref{abVrelation}), one of the matrix elements $%
V_{1\left( 2\right) }$ equals zero. For instance, the effective second-order
equation for the upper spinor component takes the simple form: 
\begin{equation}
\frac{d^{2}f_{j}}{dr^{2}}+\left[ E^{2}-m^{2}-\frac{j\left( j-1\right) }{r^{2}%
}+\left( E+m\right) V_{1}\left( r\right) \right] f_{j}=0.  \label{upper}
\end{equation}

This equation obviously does not contain a resonance state because of
absence of terms of the type characteristic for the relativistic resonances $%
V^{2},$ $dV/dr$ etc \cite{zeld}.

\section{Zero-gap graphene}

Let us consider Eq. (\ref{charbesselHJ}) in the zero-gap case $m=0$. The
symmetry Eq. (\ref{symm}) allows us to restrict the analysis by the right
energy half-plane. If we consider explicitly only the right energy
half-plane $\left( \alpha =1\right) $, we should take into account both
positive and negative angular momentum quantum numbers $j=\pm 1/2,\pm
3/2,\ldots $and both positive and negative potential amplitude $a$ values in
order to obtain a complete picture, while considering both $\alpha =1$ and $%
\alpha =2$ cases it was enough to take into account only positive $j$. The
characteristic equation can be essentially simplified in the zero-gap limit $%
m=0:$

\begin{equation*}
E\left\{ J_{j-1/2}\left( Er_{0}\right) H_{j+1/2}^{\left( 1\right) }\left(
Er_{0}\right) +J_{j+1/2}\left( Er_{0}\right) H_{j-1/2}^{\left( 1\right)
}\left( Er_{0}\right) \right. +
\end{equation*}

\begin{equation}
-T\left( a,b\right) \left. \left[ \left( a+b\right) J_{j-1/2}\left(
Er_{0}\right) H_{j-1/2}^{\left( 1\right) }\left( Er_{0}\right) -\left(
a-b\right) J_{j+1/2}\left( Er_{0}\right) H_{j+1/2}^{\left( 1\right) }\left(
Er_{0}\right) \right] \right\} =0,  \label{chareqzerogap}
\end{equation}

\begin{equation*}
j=\pm 1/2,\pm 3/2,\cdots ,\text{ \ \ }\func{Re}E\geq 0.
\end{equation*}%
Notice that this equation has a root $E=0.$ Vanishing of the energy
imaginary part stems obviously from vanishing of the free Dirac density of
states $g\left( E\right) =Tr\func{Im}G_{r}\left( E\right) $ at $E=0.$ Here $%
G_{r}\left( E\right) $ is the free retard Green function.

Complex roots of the characteristic equation will be interpreted here as
resonance states. Now we will study a distribution of the complex roots of
Eq. (\ref{chareqzerogap}), i. e. resonances.

It is convenient to divide the plane $\left( a,b\right) $ into four regions
separated by the coordinate angles bisectrices $a^{2}=b^{2}$ (see Fig. 1)$.$
We numerate these regions from I to IV.

\subsection{Resonances: $a^{2}>b^{2}$}

Let us consider at first the region I of the $\left( a,b\right) $-plane (see
Fig. 1). It is useful to introduce there the hyperbolic variables$:$%
\begin{equation}
a=\rho _{1}\cosh \psi _{1},\text{ \ \ }b=\rho _{1}\sinh \psi _{1}.
\label{hyperbol}
\end{equation}%
Then Eq. (\ref{chareqzerogap}) takes the form:%
\begin{align}
& J_{j-1/2}\left( Er_{0}\right) H_{j+1/2}^{\left( 1\right) }\left(
Er_{0}\right) +J_{j+1/2}\left( Er_{0}\right) H_{j-1/2}^{\left( 1\right)
}\left( Er_{0}\right)  \notag \\
& =\tan \rho _{1}\left[ \exp \left( \psi _{1}\right) J_{j-1/2}\left(
Er_{0}\right) H_{j-1/2}^{\left( 1\right) }\left( Er_{0}\right) -\exp \left(
-\psi _{1}\right) J_{j+1/2}\left( Er_{0}\right) H_{j+1/2}^{\left( 1\right)
}\left( Er_{0}\right) \right] .  \label{chareqa>b}
\end{align}

Periodicity of tan$\rho _{1}$ allows us to restrict the analysis here by the
segment $0\leq \rho _{1}\leq \pi .$ Consideration of only right half-plane
of energies $\func{Re}$ $E>0$ and $j=\pm 1/2,\pm 3/2,\ldots $will be suffice
due to the symmetry Eq. (\ref{symm})$.$

For a fixed $\rho _{1},$ we can consider the asymptotic behavior at $%
\left\vert \psi _{1}\right\vert \rightarrow \infty .$ Eq. (\ref{chareqa>b})
asymptotically transforms into the equations 
\begin{equation}
J_{j-1/2}\left( Er_{0}\right) H_{j-1/2}^{\left( 1\right) }\left(
Er_{0}\right) =0,\text{ \ \ \ }\psi _{1}\rightarrow \infty ,
\label{legminus}
\end{equation}

\begin{equation}
J_{j+1/2}\left( Er_{0}\right) H_{j+1/2}^{\left( 1\right) }\left(
Er_{0}\right) =0,\text{ \ \ \ }\psi _{1}\rightarrow -\infty .
\label{legplus}
\end{equation}%
Since the Hankel functions $H_{j\mp 1/2}^{\left( 1\right) }\left(
Er_{0}\right) $ have no roots in the right halfplane \cite{stegun},
solutions $E_{n}^{\left( j\mp 1/2\right) }$of Eq. (\ref{legminus}), Eq. (\ref%
{legplus}) read

\begin{equation}
E_{n}^{\left( j-1/2\right) }=\lambda _{n}^{\left( j-1/2\right) }/r_{0},\text{
\ \ \ }\psi _{1}\rightarrow \infty ,\text{ \ \ \ }n=1,2,\ldots
\label{rootminus}
\end{equation}

\begin{equation}
E_{n}^{\left( j+1/2\right) }=\lambda _{n}^{\left( j+1/2\right) }/r_{0},\text{
\ \ \ \ }\psi _{1}\rightarrow -\infty ,\text{ \ \ }n=1,2,\ldots
\label{rootplus}
\end{equation}%
with $\lambda _{n}^{\left( j\mp 1/2\right) }$ being the roots of the Bessel
functions $J_{j\mp 1/2}\left( z\right) ,$ which are real and tend to
infinity. Thus we conclude that solutions of Eq. (\ref{chareqa>b}) are
distributed on the arcuated curves beared on the real axis in the points $%
E_{n}^{\left( j\mp 1/2\right) }$ (see Fig. 2)\ 

Notice that the case of $\left\vert \psi _{1}\right\vert \rightarrow \infty $
corresponds to $a^{2}\leftrightarrows b^{2}$ (see section II, Eq. (\ref%
{upper})$.$ When we move along the non-singular hyperbolae $\rho _{1}\in
\left( 0,\pi /2\right) $ from $\psi _{1}=-\infty $ to $\psi _{1}=+\infty ,$
the $n$-th arc-like curves are circumscribed in the complex plane of $E$
from the lying on the real axis point $E_{n}^{\left( j+1/2\right)
}=r_{0}^{-1}\lambda _{n}^{\left( j+1/2\right) }$to the similar point $%
E_{n}^{\left( j-1/2\right) }=r_{0}^{-1}\lambda _{n}^{\left( j-1/2\right) }.$%
\ When $\rho _{1}$ approaches $\pi /2$ (singular hyperbola)$,$ the "arc"
height tends to infinity. Moving along the arc-like curve takes place
anti-clock-wise, when $\psi _{1}$ varies from $-\infty $ to $+\infty $. A
transition between the points $E_{n}^{\left( j\pm 1/2\right) }$ corresponds
to swapping of the upper and lower components of the wave function spinor
amplitudes. However, crossing the singular hyperbolae $\pi /2$ leads to an
essential rearrangement of the arc structure. Notably, the function $\tan
\rho _{1}$ changes its sign and alternative combinations of nearest neighbor
Bessel's function roots are connected by arcs in order to satisfy the
characteristic equation. Thus, the resonance arc-like curve begins in the
point $E_{n}^{\left( j+1/2\right) }$ and moves clock-wise into the point $%
E_{n+1}^{\left( j-1/2\right) }$, when $\pi /2<\rho _{1}<\pi .$ A schematic
picture of such transitions of arcs is presented in Fig. 3. Results of
numerical calculations for $\rho _{1}\in \left( 0,\pi \right) $ presented in
Fig. 4 for $\rho _{1}\in \left( 0,\pi /2\right) $ and in Fig. 5 for $\rho
_{1}\in \left( \pi /2,\pi \right) .$ They are in good qualitative agreement
with our analytic consideration. A periodicity with the period $\Delta \rho
_{1}=\pi $ takes place.

Motion along rays in the region I of the $\left( a,b\right) $-plane maps
into closed resonance trajectories in the energy complex plane in the case
of the ray lying in the upper half-plane. In the case of the ray lying in
the lower half-plane we obtain repeating jumps from one resonance to the
next one instead of the closed curves (see Fig. 6)..When the ray approaches
the coordinate angle bisectrix, the closed curves diameter decreases so that
the curves shrink into a point.

Consider now the left quarter (region III in Fig. 1):\ 
\begin{equation}
a=-\rho _{3}\cosh \psi _{3},\text{ }b=-\rho _{3}\sinh \psi _{3}.
\label{hyperbol3}
\end{equation}%
Then the characteristic equation reads%
\begin{align}
& J_{j-1/2}\left( r_{0}E\right) H_{j+1/2}^{\left( 1\right) }\left(
r_{0}E\right) +J_{j+1/2}\left( r_{0}E\right) H_{j-1/2}^{\left( 1\right)
}\left( r_{0}E\right)  \notag \\
& =\tan \rho _{3}\left[ \exp \left( -\psi _{3}\right) J_{j+1/2}\left(
r_{0}E\right) H_{j+1/2}^{\left( 1\right) }\left( r_{0}E\right) -\exp \left(
\psi _{3}\right) J_{j-1/2}\left( r_{0}E\right) H_{j-1/2}^{\left( 1\right)
}\left( r_{0}E\right) \right] .  \label{chaeq3}
\end{align}%
The asymptotic form of the equation at $\left\vert \psi _{3}\right\vert
\longrightarrow \infty $ is following%
\begin{equation*}
J_{j-1/2}\left( r_{0}E\right) =0,\text{ }\psi _{3}\rightarrow \infty ,
\end{equation*}

\begin{equation*}
J_{j+1/2}\left( r_{0}E\right) =0,\text{ }\psi _{3}\rightarrow -\infty .
\end{equation*}%
When $\psi _{3}$ varies from $-\infty $ to $+\infty $ circumscribing the
hyperbola in the region III, we have motion from the roots of $%
J_{j+1/2}\left( r_{0}E\right) $ to the roots of $J_{j-1/2}\left(
r_{0}E\right) $ in the energy complex plane (the same arc-form curves
circumscribed anti-clock-wise or clock-wise respectively for $\rho _{3}\in
(0,$ $\pi /2)$ or $\left( \pi /2,\pi \right) $). Numeric calculations are
similar to the case of the region I; they confirm conclusions of our
analysis.

An increase of the potential radius $r_{0}$ shifts the roots in the
direction of the coordinate origin.

\subsection{Resonances: $b^{2}>a^{2}$}

Now we consider the region II in the $\left( a,b\right) $-plane (see Fig.
1). We introduce the hyperbolic variables here differently:

\begin{equation}
b=\rho _{2}\cosh \psi _{2},\text{ }a=\rho _{2}\sinh \psi _{2}.
\label{hyperbol2}
\end{equation}%
The characteristic equation takes the form%
\begin{align}
& J_{j-1/2}\left( r_{0}E\right) H_{j+1/2}^{\left( 1\right) }\left(
r_{0}E\right) +J_{j+1/2}\left( r_{0}E\right) H_{j-1/2}^{\left( 1\right)
}\left( r_{0}E\right)  \notag \\
& =\left[ \exp \left( \psi _{2}\right) J_{j-1/2}\left( r_{0}E\right)
H_{j-1/2}^{\left( 1\right) }\left( r_{0}E\right) +\exp \left( -\psi
_{2}\right) J_{j+1/2}\left( r_{0}E\right) H_{j+1/2}^{\left( 1\right) }\left(
r_{0}E\right) \right] \tanh \rho .  \label{chareqb>a}
\end{align}%
One can see that in the limit of $a\leftrightarrows b$ we have the equation%
\begin{equation*}
J_{j-1/2}\left( r_{0}E\right) =0,\text{ \ \ \ }\psi _{2}\rightarrow \infty ,
\end{equation*}%
while for $a\leftrightarrows -b$ we have%
\begin{equation*}
J_{j+1/2}\left( r_{0}E\right) =0,\text{ \ \ \ }\psi _{2}\rightarrow -\infty .
\end{equation*}%
Thus, the roots circumscribe arc-form curves clock-wise from the point $%
\lambda _{n}^{\left( j-1/2\right) }/r_{0}$ to $\lambda _{n}^{\left(
j+1/2\right) }/r_{0}$ (see Fig. 7), while $\psi _{2}$ varies from $+\infty $
to $-\infty $ (motion along the upper hyperbolae in Fig. 1 anti-clock-wise).

Let us learn a limiting form of the arcs, when $\rho _{2}\longrightarrow
\infty .$ As $\tanh \rho _{2}\longrightarrow 1,$ Eq. (\ref{chareqb>a}) takes
the form%
\begin{equation}
\left[ \mathcal{J}_{j}\left( r_{0}E_{jn}\right) -\exp \left( -\psi
_{2}\right) \right] \left[ \mathcal{H}_{j}\left( r_{0}E_{jn}\right) -\exp
\left( -\psi _{2}\right) \right] =0,  \label{largerhoeq}
\end{equation}%
where $E_{jn}=E_{jn}^{\left( 1\right) }-i\Gamma _{jn}/2$ is the complex root
of the characteristic equation; we used the standard notions for the real
and imaginary parts of the resonance energy.

Notice that roots of the second factor are not physical and must be ignored.
Eq. (\ref{largerhoeq}) can be written in the form%
\begin{equation}
\mathcal{J}_{j}\left( r_{0}\left( E_{jn}^{\left( 1\right) }-i\Gamma
_{jn}/2\right) \right) =\exp \left( -\psi _{2}\right) ,\text{ }\rho
_{2}\longrightarrow \infty ,  \label{largerhoeq2}
\end{equation}%
Since $\mathcal{J}_{j}\left( z\right) $ is real on the real axis, we have $%
\mathcal{J}_{j}\left( z\right) ^{\ast }=\mathcal{J}_{j}\left( z^{\ast
}\right) $ and, therefore, $\mathcal{J}_{j}\left( r_{0}\left( E_{jn}^{\left(
1\right) }-i\Gamma _{jn}/2\right) \right) =\mathcal{J}_{j}\left( r_{0}\left(
E_{jn}^{\left( 1\right) }+i\Gamma _{jn}/2\right) \right) ^{\ast }=\exp
\left( -\psi _{2}\right) .$ So we see that $E_{jn}^{\left( 1\right)
}+i\Gamma _{jn}/2$ is a root of Eq. (\ref{largerhoeq2}) as well. Since the
ends of the arc are simple zeros of the Bessel function, these two arcs
coincide and are flat in this limit. Eq. (\ref{largerhoeq2}) takes the form

\begin{equation}
\mathcal{J}_{j}\left( r_{0}E_{jn}^{\left( 1\right) }\right) \equiv
J_{j-1/2}/J_{j+1/2}=\exp \left( -\psi _{2}\right) =\sqrt{\left( b-a\right)
/\left( b+a\right) },\text{ \ \ }b^{2}\gg a^{2}.  \label{largerhoreal}
\end{equation}%
We see that the energy imaginary part is getting extremely small and the
arcs approach the real axis segments $[\lambda _{n}^{\left( j-1/2\right)
}/r_{0},$ $\lambda _{n}^{\left( j+1/2\right) }/r_{0}]$ so fast at $\rho
_{2}\longrightarrow \infty $ that it is difficult to observe a presence of
the imaginary part numerically. However, this process can be followed
analytically. In order to investigate the limiting behavior of the energy
imaginary part, we expand Eq. (\ref{chareqb>a}) for $\rho
_{2}\longrightarrow \infty $ taking account of the asymptotics leading
terms. In particular, the expansion $\tanh x\cong 1-\exp \left( -2x\right)
+\ldots $ for $x\gg 1$ was used. Taking the real part of Eq. (\ref{chareqb>a}%
), we obtain after some algebra 
\begin{eqnarray*}
&&\func{Re}\mathcal{J}_{j}(r_{0}E_{jn})\left[ 1-\left( 1-\exp (-2\rho
_{2})\exp \left( \psi _{2}\right) \func{Re}\mathcal{H}_{j}\left(
r_{0}E_{jn}\right) \right) \right] \\
&=&-\func{Re}\mathcal{H}_{j}\left( r_{0}E_{jn}\right) +\Gamma _{n}\mathcal{J}%
_{j}^{^{^{\prime }}}\left( r_{0}E_{jn}^{\left( 1\right) }\right) \left(
r_{0}/2\right) \left( 1-\exp \left( -2\rho _{2}\right) \right) \exp \left(
-\psi _{2}\right) \func{Im}\mathcal{H}_{j}\left( r_{0}E_{jn}\right) \\
&&+\left( 1-\exp \left( -2\rho _{2}\right) \right) \exp \left( -\psi
_{2}\right) .
\end{eqnarray*}

Expanding this formula for small $\Gamma _{jn}$ we obtain in the leading
order

\begin{equation}
r_{0}\Gamma _{jn}/2=-\frac{\exp \left( -\psi _{2}-2\rho _{2}\right) }{%
\mathcal{J}_{j}^{^{^{\prime }}}\left( r_{0}E_{jn}^{\left( 1\right) }\right) }%
\cdot \frac{\func{Re}\mathcal{H}_{j}\left( r_{0}E_{jn}^{\left( 1\right)
}\right) +\exp \left( -\psi _{2}\right) }{\func{Im}\mathcal{H}_{j}\left(
r_{0}E_{jn}^{\left( 1\right) }\right) }.  \label{impart}
\end{equation}%
Here $E_{jn}^{\left( 1\right) }$ is determined from Eq. (\ref{largerhoreal}%
). It is seen from Eq. (\ref{impart}) that the resonance width $\Gamma _{jn}$
decreases exponentially at large $\rho _{2}.$

Returning to the initial notions we find:

\begin{eqnarray}
r_{0}\Gamma _{jn} &=&2\exp \left( -2\sqrt{b^{2}-a^{2}}\right) \frac{\sqrt{%
\left( b-a\right) /\left( b+a\right) }J_{j+1/2}^{2}}{J_{j+1/2}^{%
%TCIMACRO{\U{b4}}%
%BeginExpansion
{\acute{}}%
%EndExpansion
}J_{j-1/2}-J_{j-1/2}^{%
%TCIMACRO{\U{b4}}%
%BeginExpansion
{\acute{}}%
%EndExpansion
}J_{j+1/2}}  \notag \\
&&\cdot \frac{\func{Re}\left[ H_{j-1/2}^{\left( 1\right) }/H_{j+1/2}^{\left(
1\right) }\right] +\sqrt{\left( b-a\right) /\left( b+a\right) }}{\func{Im}%
\left[ H_{j-1/2}^{\left( 1\right) }/H_{j+1/2}^{\left( 1\right) }\right] },%
\text{ \ \ \ \ \ \ \ \ }b^{2}>>a^{2}  \label{imeq}
\end{eqnarray}

The obtained formula indicates that these resonances are extremely sharp for
the case of small $\left\vert j\right\vert $ and, therefore, they can give a
strong resonance scattering of electrons. Using the Bessel functions $J_{\nu
},$ $H_{\nu }^{\left( 1\right) }$ asymptotics for large $\nu $ \cite{stegun}%
, we obtain the following estimate for the resonance width asymptotic at
large $\left\vert j\right\vert :$%
\begin{equation}
\Gamma _{jn}\sim \left( r_{0}^{-1}\right) \sqrt{\left( b-a\right) /\left(
b+a\right) }\exp \left( -2\sqrt{b^{2}-a^{2}}\right) \left( \left\vert
j\right\vert /e\right) ^{\left\vert j\right\vert }.  \label{largej}
\end{equation}

This function catastrophically increases beginning from $\left\vert
j\right\vert \sim 4.$ It means that the imaginary part of the energy is
large for any realistic $\rho _{2}$ so that solutions with $j>4$ are not
really sharp resonances and they will not contribute into the electron
resonance scattering.

Results of numeric calculation for resonances in the case of $b^{2}>a^{2}$
are presented in Figs. 7 and 8. Approaching of arc-like resonance curves to
the real axis segments, when $\rho _{2}$ increases, is shown in Fig. 7. Fig.
8 shows the resonance trajectory corresonding to a motion along the ray in
the $\left( a,b\right) $-plane region II. When the radius increases, the
trajectory approaches a point on the real axis.

The region IV (lower quarter) can be analyzed similarly using the hyperbolic
variables: 
\begin{equation}
b=-\rho _{4}\cosh \psi _{4},\text{ }a=-\rho _{4}\sinh \psi _{4}
\label{hyper4}
\end{equation}

An analisis of this region could be done similarly to the region II, but we
omit it here since no qualitative distinction is observed in this case.
Numerical results for the region IV are omitted by us as well because of
this similarity.

So we have discussed a local behavior of the resonance trajectories within
the chosen quarter in the $\left( a,\text{ }b\right) $-plane. Now we will
consider the global behavior, when all quarters are passed along hyperbolae
with the fixed hyperbolic radius $\rho .$ This process creates pairs of
arc-form curves bearing on two common basing points lying on the real axis.
These figures of eight are arranged at the $E$ complex plane symmetrically
relative to the imaginary axis in the case of $m=0.$ Resonances move
anti-clock-wise from the root $\lambda _{n}^{\left( j+1/2\right) }/r_{0}$ to
the root $\lambda _{n}^{\left( j-1/2\right) }/r_{0}$ circumscribing the arc,
when $\phi \in \left[ -\pi /4,\text{ }\pi /4\right] $ When $\phi \in \left[
\pi /4,\text{ }3\pi /4\right] ,$ a different arc appears. It connects the
same points $\lambda _{n}^{\left( j+1/2\right) }$ and $\lambda _{n}^{\left(
j-1/2\right) }$, but moving in the opposite direction (clock-wise). We
obtain in result the figures-of-eight in the $E$-complex plane. This picture
is repeated with the period $\Delta \phi =\pi $.

Such excursion around all quarters of the $\left( a,\text{ }b\right) $-plane
can be carried out circumscribing the circles in the $\left( a,b\right) $%
-plane%
\begin{equation}
a=\rho \cos \phi ,\text{ }b=\rho \sin \phi .  \label{circle}
\end{equation}%
One quarter after another is passed in this case too, but now the
figures-of-eight do not touch the real energy axis anymore (see Fig. 9), i.
e. all resonances belonging to this set have rather large widths. Numeric
analysis confirms our prediction on the mapping of the circular excursion in
the $\left( a,b\right) $ plane onto the energy plane. When the perturbation
intensity $\rho $ is increased, the picture is getting more complicated: all
the figures-of-eight are associating into a single whole (see Fig. 9 for
large $\rho $). Notice that the resonance trajectory self-crossing points
are fixed points relative to $\rho $ changes. {\LARGE !}They correspond to
resonance set, which do not change at motion along the ray $a=b.$

\section{Gapped graphene}

While the pristine graphene is gapless, a contact with the substrate can
induce some narrow gap \cite{gap}. That is why we study here bound and
resonance states in the gapped graphene as well.

Let us consider the characteristic equation for the case of $m\neq 0.$ We
introduce\ the dimensionless parameters:%
\begin{equation}
\epsilon =E/m,\text{ }r_{0}m=r_{0}/r_{c}=\gamma ,  \label{dimensionless}
\end{equation}%
where $r_{c}=1/m$ is the "Compton radius" ($\hslash =1,$ $v_{F}=1).$ The
characteristic equation for bound states Eq. (\ref{character1a}) takes the
form:%
\begin{eqnarray}
&&\sqrt{1-\epsilon ^{2}}\left[ I_{j-1/2}\left( z\right) K_{j+1/2}\left(
z\right) -K_{j-1/2}\left( z\right) I_{j+1/2}\left( z\right) \right]  \notag
\\
&=&T(a,b)\left[ \left( 1-\epsilon \right) \left( a-b\right) J_{j+1/2}\left(
z\right) K_{j+1/2}\left( z\right) +\left( 1+\epsilon \right) \left(
a+b\right) I_{j-1/2}\left( z\right) K_{j-1/2}\left( z\right) \right] ,
\label{massiveeq}
\end{eqnarray}%
where $z=\gamma \sqrt{1-\epsilon ^{2}}.$ The perturbation radius $r_{0}$
value is of order of the lattice spacing $d,$ while the "Compton" radius $%
r_{c}$ is rather large: $r_{c}>>d:$ $\hslash /\left( m_{0}v_{F}\right) \sim
10^{-8}$ $cm$ even for the bare electron mass $m_{0}$; the real effective
mass at the band extrema can be estimated as $10^{-1}-10^{-2}$ of $m_{0}$.
Therefore, the actual value of $\gamma $ is less than unity.

We consider here some general properties of the bound states electronic
spectrum resulting from the characteristic equation (\ref{massiveeq}) and
the effect of non-vanishing gap on the resonance states. Results of the
numerical solution will be\ presented too.

\subsection{Bound states: $a^{2}>b^{2}$}

Let us assume at first $j=1/2.$ We consider Eq. (\ref{massiveeq}) for the
energy situated in the gap $\epsilon ^{2}<1$. We use the Bessel functions
limiting forms for the states lying near the gap edges \cite{stegun}

\begin{equation}
I_{\nu }(z)\sim \left( z/2\right) ^{\nu }\frac{1}{\Gamma \left( \nu
+1\right) },\text{ }K_{0}\left( z\right) \sim -\log z,\text{ }K_{\nu }\left(
z\right) \sim \frac{1}{2}\Gamma \left( \nu \right) \left( z/2\right) ^{-\nu
},  \label{expansion}
\end{equation}%
in order to transform the characteristic equation (\ref{massiveeq}). Here $%
\Gamma \left( \nu \right) $ is the gamma function. We obtain a simple
relation for small $z.$

\begin{eqnarray}
&&1+\left( \gamma ^{2}/4\right) \left( 1-\epsilon ^{2}\right) \log \left(
\gamma \sqrt{1-\epsilon ^{2}}\right)  \notag \\
&\cong &\gamma \frac{\tan \sqrt{a^{2}-b^{2}}}{\sqrt{a^{2}-b^{2}}}\left[
\left( 1-\epsilon \right) \left( a-b\right) /4-\left( 1+\epsilon \right)
\left( a+b\right) \log \left( \gamma \sqrt{1-\epsilon ^{2}}\right) \right] .
\label{approx}
\end{eqnarray}%
Formulae like Eq. (\ref{approx}) determine a map from the plane $\left(
a,b\right) $ onto the complex energy plane; in the case of bound states the
map is carried out onto the real energy axis segment $-1<\epsilon <1.$Now we
consider the asymptotic energy level behavior near the upper edge of the gap 
$\epsilon \longrightarrow 1.$ Eq. (\ref{approx}) takes the asymptotic form:

\begin{equation}
1\cong -\gamma \frac{\tan \sqrt{a^{2}-b^{2}}}{\sqrt{a^{2}-b^{2}}}\left(
a+b\right) \log \left[ 2\gamma ^{2}\left( 1-\epsilon \right) \right] .
\label{asymptot}
\end{equation}

Therefore, we can write

\begin{equation}
1-\epsilon \cong \left( 2\gamma ^{2}\right) ^{-1}\exp \left[ -\left(
1/\gamma \right) \frac{1}{a+b}\cdot \frac{\sqrt{a^{2}-b^{2}}}{\tan \sqrt{%
a^{2}-b^{2}}}\right] .  \label{upperedge}
\end{equation}

It is clear that $\epsilon \longrightarrow 1$ \ if $a+b\longrightarrow +.$ $%
0.$ This result conforms the well known general property of the
two-dimensional quantum systems: a threshold for creation of the bound state
is absent for $j=1/2$.

Formula (\ref{upperedge}) can be rewritten in the hyperbolic variables in
the region I (see Eq. (\ref{hyperbol}) as follows

\begin{equation}
1-\epsilon \cong \left( 2\gamma ^{2}\right) ^{-1}\exp \left[ -1/\left(
\gamma \exp \left( \psi _{1}\right) \tan \rho _{1}\right) \right] .
\label{upperedgeA}
\end{equation}%
It is seen from the last formula that $\epsilon \longrightarrow 1$, when $%
\rho _{1}\longrightarrow \pi n+0$ (where $n=1,$ $2\ldots )$, i. e. the $%
\epsilon =1$ "fronts" approach the hyperbolae $\rho _{1}=\sqrt{a^{2}-b^{2}}%
\longrightarrow \pi n+0.$

Let us study now a behavior of the bound state trajectory near the lower
band edge. Using Eq. (\ref{approx}) we conclude that when $\epsilon
\longrightarrow -1,$ the following relation holds:

\begin{equation}
1\cong \gamma \tan \left( \rho _{1}\right) \exp \left( -\psi _{1}\right)
\left( 1-\epsilon \right) /4.  \label{loweredge}
\end{equation}%
This equation shows, how the bound states trajectory approaches the lower
edge of the gap. Putting $\epsilon =-1$ we obtain the equation, which
determines the $\epsilon =-1$ boundary lines in the $\left( a,b\right) $%
-plane, separating the domains of the bound states:%
\begin{equation}
\tan \rho _{1}=2\exp \psi _{1}/\gamma .  \label{rightfront}
\end{equation}%
When $\psi _{1}\longrightarrow +\infty $ along one of these lines, $\rho
_{1}\longrightarrow \pi /2+\pi n-0,$ $n=0,1,\ldots $. The line $n=0$ crosses
the abscissa axis in the point (see Eq. (\ref{loweredge})): 
%\begin{subequations}
\begin{equation}
b_{0}=0,\text{ }a_{0}=\arctan \left( 2/\gamma \right) .  \label{ascissacross}
\end{equation}%
Then this curve crosses the bisectrix $a+b=0$ in the point $a_{1}=1/\gamma ,$
$b_{1}=-1/\gamma .$ Any point situated in the domain restricted at the left
by the bisectrix $a-b=0$ and at the right by the line Eq. (\ref{rightfront})
($\epsilon =-1$ "front") (which asymptotically approaches the hyperbola $%
\rho _{1}=\pi /2-0,$ $\psi _{1}\longrightarrow +\infty $), gives a solution
of the characteristic equation lying in the gap $\left\vert \epsilon
\right\vert <1$ that is an eigenvalue of our problem. A motion from the
bisectrix $a+b=0$ to the curve Eq. (\ref{rightfront}) in the region I of the 
$(a,b)$-plane is mapped to the motion from the upper to lower edges in the
energy gap.

Other domains with $n=1,2,\ldots ,$ lie between approaching the hyperbolae $%
\rho _{1}=\pi n-0$ curves ($\epsilon =1$ "front", $\left\vert \psi
_{1}\right\vert \longrightarrow \infty $) and the $\epsilon =-1$ "front",
approaching the hyperbolae $\rho _{1}=\pi /2+\pi n-0$ at $\psi
_{1}\longrightarrow +\infty $ and the hyperbolae $\rho _{1}=\pi n+0$ at $%
\psi _{1}\longrightarrow -\infty .$ These lines boarder the countable set of
domains in the $\left( a,b\right) $-plane, where the bound states exist.
When we move along the lying in the region I ray beginning in the coordinate
origin in the direction of the $\rho _{1}^{2}=a^{2}-b^{2}$ increase$,$ we
cross this set of domains so that every time the bound states trajectory
starts at $\epsilon =1$ and terminates at $\epsilon =-1.$

A dependence of the bound state energy on the radius $\rho _{1},$ at the
motion along the rays in the $\left( a,b\right) $-plane (see inset to Fig.
10) was investigated numerically in the case of $a^{2}>b^{2}$. The motion
between the coordinate origin and the first $\epsilon =-1$ "front" maps into
the mildly sloping \ transition between the $\epsilon =1$ and $\epsilon =-1$
edges of the gap (solid line) and the sharp sloping transition respectively
for the rays in the upper and lower half-planes of the $\left( a,b\right) $%
-plane (first pair of the curves). The second and so on pairs of curves show
similar transitions between the domain boundaries.

Now we consider the region III. Let us return to Eq. (\ref{approx}) that is
valid near the band edges and everywhere inside the gap if $\gamma <<1.$
Using the hyperbolic variables Eq. (\ref{hyperbol3}) we can write for the
bound states near the upper edge $\epsilon =+1:$

%\end{subequations}
\begin{equation}
1-\epsilon \cong \left( 2\gamma ^{2}\right) ^{-1}\exp \left[ \frac{\exp
\left( \psi _{3}\right) }{\gamma \tan \rho _{3}}\right] .  \label{upper3}
\end{equation}%
Eq. (\ref{upper3}) is satisfied near the hyperbolae $\rho _{3}=\pi n-0$
(then $\tan \rho _{3}\longrightarrow -0).$ This determines the $\epsilon =1$
"front" of domains containing the eigenvalues. The $\epsilon =1$ "fronts"
are formed by the approaching the hyperbolae $\rho _{3}=\pi n-0$, ($%
n=1,2,\ldots )$ at $\left\vert \psi _{3}\right\vert \longrightarrow +\infty
. $ Therefore, we have the exponential approaching of the energy level to
the upper band edge in the region III similarly to the region I.

The $\epsilon =-1$ "fronts" Eq. (\ref{rightfront})%
\begin{equation*}
\tan \rho _{3}=-\left( 2/\gamma \right) \exp \psi _{3}
\end{equation*}%
approach the hyperbolae $\rho _{3}=\pi n-0,$ when $\psi _{3}\longrightarrow
-\infty ,$ and hyperbolae $\rho _{3}=-\pi /2+\pi n+0,$ when $\psi
_{3}\longrightarrow +\infty ,$ $n=1,2,\ldots .$

\subsection{Bound states: $b^{2}>a^{2}$}

Let us consider the region II in the $\left( a,b\right) $-plane.The
approximate characteristic equation Eq. (\ref{approx}) for the states near
the band edges in the case of $b^{2}>a^{2}$ reads:

\begin{eqnarray}
&&1+\left( \gamma ^{2}/4\right) \left( 1-\epsilon ^{2}\right) \log \left(
\gamma \sqrt{1-\epsilon ^{2}}\right)  \notag \\
&\cong &\gamma \frac{\tanh \sqrt{b^{2}-a^{2}}}{\sqrt{b^{2}-a^{2}}}\left[
\left( 1-\epsilon \right) \left( a-b\right) /4-\left( 1+\epsilon \right)
\left( a+b\right) \log \left( \gamma \sqrt{1-\epsilon ^{2}}\right) \right] .
\label{b>aeq}
\end{eqnarray}

We obtain for the energy near the upper band gap edge:

\begin{equation*}
1-\epsilon \cong \left( 2\gamma ^{2}\right) ^{-1}\exp \left[ -\left(
1/\gamma \right) \frac{1}{a+b}\cdot \frac{\sqrt{b^{2}-a^{2}}}{\tanh \sqrt{%
b^{2}-a^{2}}}\right] .
\end{equation*}

Thus, the $\epsilon =1$ "front" approaches the bisectrix $a+b\longrightarrow
0$ exponentially. We introduce here the variables 
\begin{equation}
b=\rho _{2}\cosh \psi _{2},\text{ }a=\rho _{2}\sinh \psi _{2}.
\label{hyper2}
\end{equation}%
Then Eq. (\ref{b>aeq}) takes the form

%\begin{subequations}
\begin{eqnarray}
&&1+\left( \gamma ^{2}/4\right) \left( 1-\epsilon ^{2}\right) \log \left(
\gamma \sqrt{1-\epsilon ^{2}}\right)  \notag \\
&\cong &-\gamma \tanh \rho _{2}\left[ \left( 1-\epsilon \right) \exp \left(
-\psi _{2}\right) /4+\left( 1+\epsilon \right) \exp \left( \psi _{2}\right)
\log \left( \gamma \sqrt{1-\epsilon ^{2}}\right) \right] .  \label{b>a2}
\end{eqnarray}%
We obtain for the energies near the upper band edge $\epsilon
\longrightarrow 1$:

%\end{subequations}
\begin{equation}
1-\epsilon \cong 1/\left( 2\gamma ^{2}\right) \exp \left[ -\frac{\exp \left(
-\psi _{2}\right) }{\gamma \tanh \rho _{2}}\right]  \label{asymptotic}
\end{equation}%
Thus, the $\epsilon =1$ "front" approaches the bisectrix $a+b=0$ at $\rho
_{2}\longrightarrow 0.$ There is no "front" at the bisectrix $a-b=0$ as it
was shown above. Eq. (\ref{b>a2}) can be rewritten for $\rho
_{2}\longrightarrow \infty :$

\begin{equation}
\left[ 1+\left( \gamma /4\right) \left( 1-\epsilon \right) \exp \left( -\psi
_{2}\right) \right] \left[ 1+\gamma \left( 1+\epsilon \right) \exp \left(
\psi _{2}\right) \log \left( \gamma \sqrt{1-\epsilon ^{2}}\right) \right]
\cong 0.  \label{asimptotic2}
\end{equation}%
Therefore,%
\begin{equation}
\gamma \left( 1+\epsilon \right) \exp \left( \psi _{2}\right) \log \left(
\gamma \sqrt{1-\epsilon ^{2}}\right) \cong -1.  \label{-1}
\end{equation}%
Eq. (\ref{-1}) is satisfied if $\epsilon \longrightarrow -1$ and $\psi
_{2}\longrightarrow +\infty .$ Therefore, the $\epsilon =-1$ "front"
approaches the bisectrix $a-b=0,$ if $\rho _{2}\longrightarrow \infty .$This
"front" lies in the region I (see above). Thus, a motion in the region II of
the plane $\left( a,b\right) $ from the bisectrix $a+b=0$ to the bisectrix $%
a-b=0$ corresponds to a motion of the energy eigenvalue from the upper band
edge $\epsilon =1$ to the lower band edge $\epsilon =-1$, when $\rho
_{2}>>1. $

A dependence of the bound state energy on the radius $\rho _{2},$ at the
motion along the rays in the $\left( a,b\right) $-plane (see inset to Fig.
11) was investigated numerically in the case of $a^{2}<b^{2}$. The motion
along the ray maps into the monotone transition of the energy level from the
band gap edge $\epsilon =1$ to the constant situated in the gap. Motion
along the ray in the right half-plane maps into the faster energy dependence
and with lower asymptotic energy value, than in the case of the ray in the
left half-plane. When the ray approaches the bisectrix $a-b=0$, the
asymptotic value tends to $\epsilon =-1.$

Let us consider now solutions of the characteristic equation near the band
edges for the parameters $a,b$ in the region IV. We introduce the hyperbolic
variables (see Eq. (\ref{hyper4})).

Then Eq. (\ref{b>aeq}) can be approximately written in the form:

\begin{eqnarray}
&&1+\left( \gamma ^{2}/4\right) \left( 1-\epsilon ^{2}\right) \log \left(
\gamma \sqrt{1-\epsilon ^{2}}\right)  \notag \\
&\cong &\gamma \tanh \rho _{4}\left[ \left( 1-\epsilon \right) \exp \left(
-\psi _{4}\right) /4+\left( 1+\epsilon \right) \exp \left( +\psi _{4}\right)
\log \left( \gamma \sqrt{1-\epsilon ^{2}}\right) \right] .  \label{edge4}
\end{eqnarray}%
The line%
\begin{equation}
1\cong \left( \gamma /2\right) \exp \left( -\psi _{4}\right) \tanh \rho _{4}
\label{front4}
\end{equation}%
corresponds to the $\epsilon =-1$ "front" in the region IV. This line is a
continuation of the line Eq. (\ref{rightfront}); they meet in the point $%
a=-b=1/\gamma .$ The line Eq. (\ref{front4}) turns into the asymptotic at $%
\rho _{4}\longrightarrow \infty :$%
\begin{equation}
\psi _{4}\longrightarrow -\log \left( 2/\gamma \right) .  \label{psi4asympt}
\end{equation}

Let us transform Eq. (\ref{edge4}) in the limit of $\rho _{4}\longrightarrow
\infty :$%
\begin{equation}
\left[ 1-\left( \gamma /4\right) \left( 1-\epsilon \right) \exp \left( -\psi
_{4}\right) \right] \left[ 1-\gamma \left( 1+\epsilon \right) \exp \left(
\psi _{4}\right) \log \left( \gamma \sqrt{1-\epsilon ^{2}}\right) \right]
\cong 0.  \label{edge4a}
\end{equation}

The "front" $\epsilon \longrightarrow 1$ equation takes the form 
\begin{equation}
1-\epsilon \cong \left( 4/\gamma \right) \exp \left( \psi _{4}\right)
\label{1stfront}
\end{equation}

Thus, the "front" $\epsilon \longrightarrow 1$ approaches the bisectrix $%
b+a=0$ asymptotically, when $\psi _{4}\longrightarrow -\infty .$ Therefore,
a motion from the line Eq. (\ref{psi4asympt}) to the bisectrix $b+a=0$ in
the $\left( a,b\right) $ plane is mapped to a motion from the lower edge of
the gap to the upper one when $\rho _{4}>>1.$

If $\gamma $ increases$,$ we conclude from Eq. (\ref{psi4asympt}) that the
second front $1+\epsilon \longrightarrow 0$ moves to the bisectrix $b-a=0,$
i. e. an increase of the mass widens the domain of the region IV, where the
eigenvalues exist; it fills all region IV in the extreme limit.of large $%
\gamma .$ Notice that this case is not realistic for an impurity problem,
but can be important for the quantum dot case \cite{matulis}.

When $\gamma <<1,$ it is seen from Eq. (\ref{psi4asympt}) that the "front" $%
1+\epsilon \longrightarrow 0$ tends to the "front" $1-\epsilon
\longrightarrow 0$ at $\psi _{4}\longrightarrow -\infty ,$ i. e. the fronts
tend to merge, when the gap tends to zero.

Thus, our analysis shows that there is a countable set of eigenvalue domains
in the case $a^{2}>b^{2}.$ The boundaries of these domains approach the
hyperbolae described above. In the case of $b^{2}>a^{2},$ we have a
saturation instead of periodicity, when radius increases.

\subsection{Bound states for higher angular momenta}

Let us consider now the eigenvalue spectrum for the angular momentum $j=3/2,$%
for instance, when the parameters $a,$ $b$ lie in the regions I and II..
Using the Bessel functions expansion for small arguments (\ref{expansion}),
Eq. (\ref{massiveeq}) can be written as follows:%
\begin{equation}
1/\gamma -\gamma \left( 1-\epsilon ^{2}\right) /12=T\left( a,b\right) \left[
\left( a-b\right) \left( 1-\epsilon \right) /12+\left( a+b\right) \left(
1+\epsilon \right) /4\right] .  \label{3/2eq}
\end{equation}%
In the region I, this equation takes the form

\begin{equation}
1/\gamma -\gamma \left( 1-\epsilon ^{2}\right) /12=\tan \rho _{1}\left[
\left( 1-\epsilon \right) \exp \left( -\psi _{1}\right) /12+\left(
1+\epsilon \right) \exp \left( \psi _{1}\right) /4\right] .  \label{3/2eqI}
\end{equation}

The "fronts" $\epsilon \longrightarrow 1$ have the form%
\begin{equation}
\tan \rho _{1}=\left( 2/\gamma \right) \exp \left( -\psi _{1}\right) .
\label{firstfrontI}
\end{equation}

We have asymptotically:%
\begin{equation}
\rho _{1}\left( \psi _{1}\right) \longrightarrow \left\{ 
\begin{array}{c}
\pi /2+\pi n-0\text{ \ \ \ \ for\ }\psi _{1}\longrightarrow -\infty , \\ 
\\ 
\pi n+0\text{ \ \ \ for }\psi _{1}\longrightarrow \infty ,%
\end{array}%
\text{ \ \ }n=1,2,\ldots \right.  \label{asymptotlinesI}
\end{equation}

The "fronts" $\epsilon \longrightarrow -1$ have the form%
\begin{equation}
\tan \rho _{1}=\left( 6/\gamma \right) \exp \left( \psi _{1}\right) .
\label{secondfrontI}
\end{equation}

Thus we have the lines%
\begin{equation}
\rho _{1}\left( \psi _{1}\right) \longrightarrow \left\{ 
\begin{array}{c}
\pi n+0\text{ \ \ \ \ for\ }\psi _{1}\longrightarrow -\infty , \\ 
\\ 
\pi /2+\pi n-0\text{ \ \ \ for }\psi _{1}\longrightarrow \infty .%
\end{array}%
\text{ \ \ }n=1,2,\ldots \right.  \label{asymptotsecondI}
\end{equation}

The "fronts" Eq. (\ref{firstfrontI}) ($\epsilon \longrightarrow 1)$, Eq. (%
\ref{secondfrontI}) $\left( \epsilon \longrightarrow -1\right) $ cross for
the first time the abscissa axis respectively in the points 
\begin{eqnarray}
a_{1} &=&\arctan \left( 2/\gamma \right) ,\text{ \ \ }b_{1}=0  \label{a1} \\
a_{2} &=&\arctan \left( 6/\gamma \right) .  \label{a2}
\end{eqnarray}%
Therefore, a motion from the point $a_{1}$ to the point $a_{2}$ maps into a
motion from $\epsilon =1$ to $\epsilon =-1.$ More than that, having in mind
a continuous dependence of solutions on the parameters $a$ and $b$ the
"fronts" Eq. (\ref{firstfrontI}), Eq. (\ref{secondfrontI}) obtained from the
asymptotic equation (\ref{3/2eqI}) cross one another. This prediction is
confirmed by the numerical calculations (see Fig. 13). Exclusively fast
dependence of the energy eigenvalues on the parameters $a$ and $b$ will be
observed in the vicinity of the crossing point.

Let us consider now the region II. We have here from Eq. (\ref{3/2eq}): 
\begin{equation}
1/\gamma -\gamma \left( 1-\epsilon ^{2}\right) /12=\tanh \rho _{2}\left[
-\left( 1-\epsilon \right) \exp \left( -\psi _{1}\right) /12+\left(
1+\epsilon \right) \exp \left( \psi _{1}\right) /4\right] .  \label{3/2eqII}
\end{equation}

The "front" $\epsilon \longrightarrow 1$ is given by the equation%
\begin{equation}
\tanh \rho _{2}=\left( 2/\gamma \right) \exp \left( -\psi _{2}\right) .
\label{firstfrontII}
\end{equation}

Since $0\leq \tanh \rho _{2}\leq 1$ \ $\left( \rho _{2}\geq 0\right) ,$ we
have%
\begin{equation}
\psi _{2}\geq -\log \left( \gamma /2\right) .  \label{psi2>}
\end{equation}

The domain, where the energy eigenvalues exist in the case of $\gamma =2,$
is bounded by the positive half-axis $b$ and the first quadrant bisectrix.
When $\gamma <2$ and decreases$,$ the "front" $\epsilon =1$ turns about the
coordinate origin from the ordinate axis approaching the first quadrant
bisectrix. The eigenvalues domain is narrowing in result. When $\gamma >2$
and increases, this boundary turns anti-clock-wise approaching the second
quadrant bisectrix. When $\psi _{2}\longrightarrow \infty ,$ Eq. (\ref%
{3/2eqII}) can be satisfied only if 
\begin{equation}
1+\epsilon =4/\left( \gamma \tanh \rho _{2}\right) \exp \left( -\psi
_{2}\right) ,\text{ \ \ }\psi _{2}\longrightarrow \infty  \label{-mII}
\end{equation}

We have for $\rho _{2}\longrightarrow \infty $ (higher hyperbolae):%
\begin{equation}
1+\epsilon =\left( 4/\gamma \right) \exp \left( -\psi _{2}\right) ,\text{ \
\ }\psi _{2}\longrightarrow \infty .  \label{-masymptII}
\end{equation}

When $\psi _{2}$ is fixed, the asymptotic magnitude of the eigenvalue $%
\epsilon _{as}$ at $\rho _{2}\longrightarrow \infty $ can be found from the
quadratic equation 
\begin{equation}
\epsilon _{as}^{2}-\gamma ^{-1}\left[ 6\exp \left( \psi _{2}\right) +2\exp
\left( -\psi _{2}\right) \right] \epsilon _{as}-1+12/\gamma ^{2}+\left(
2/\gamma \right) \exp \left( -\psi _{2}\right) -\left( 6/\gamma \right) \exp
\left( \psi _{2}\right) =0.  \label{quadrat}
\end{equation}

The domain of eigenvalues in the region IV is narrower in the case of $j=3/2$%
, than for $j=1/2$ similarly to the region II. Generically, an increase of $%
j $ at fixed $\gamma $ leads to narrowing of the eigenvalues existence
domains.

The case of higher angular momenta $j=5/2,\ldots $ can be analyzed
similarly. The energy discrete eigenvalues exist for all $j,$ but in
opposite to the case of $j=1/2,$ there is a threshold for the bound state
appearing at the upper edge of the gap. Existence of bound states for all
higher angular momenta is a specific feature of the delta function
perturbation.

The states with $j<0$ can be easily obtained with a use of the symmetry
transformation Eq. (\ref{symm}).

\subsection{Diagram of the bound states}

General results of mathematical and numerical analysis for a distribution of
bound states for $j=1/2$ and $j=3/2$ are presented respectively in Figs. 12,
13. Shaded domains represent basines of the bound states existence. The
thick lines correspond to the upper edge of the band gap $\epsilon =+1$,
while the thin ones represent the lower edge $\epsilon =-1.$

In the case of $j=1/2,$ domain boundaries approach asymptotically the
hyperbolae. One crossing point of boundaries can be seen. Therefore, we have
one twisted domain in this case. Notice that this crossing looks so in a
projection onto the $\left( a,b\right) $-plane. A \ three-dimensional
picture of this feature is shown.in Fig. 13: it is seen that the "fronts" $%
\epsilon =\pm 1$ lie at different "floors" so that the crossing curves in
the plane curves appear to be skew ones.

Notice that our diagrams of states are obtained assuming single-valued
dependence of the bound state energy on the parameters $a$ and $b$. This is
true everywhere in the $\left( a,b\right) $-plane except very small islands
near the domain boundaries (see Fig. 10 of this paper and our work \cite{we}%
, where the twisting point neighborhood was analyzed numerically).

In the case of $j=3/2,$the bound state existence domains are getting
narrower, a number of the boundaries crossing points increases. The diagram
of states for higher $j$ is given us in Ref. \cite{we}.

\subsection{Resonances: $m>0$}

It is necessary to return now to Eq. (\ref{charbesselHJ}) written in terms
of non-modified Bessel's functions. We consider here an effect of non-zero
bandgap upon the resonance states. It is convenient to re-write Eq. (\ref%
{charbesselHJ}) using the Bessel function explicitly (see eq (\ref{ratio2}))
and choosing $\alpha =1.$ We analyze here resonances for the parameters
values lying in the region I as an example.

Introducing the hyperbolic coordinates according to Eq. (\ref{hyperbol}) and
using the dimensionless parametrization Eq. (\ref{dimensionless}) , we can
write

\begin{eqnarray}
&&\left[ J_{j-1/2}\left( \varsigma \right) H_{j+1/2}^{\left( 1\right)
}\left( \varsigma \right) +J_{j+1/2}\left( \varsigma \right)
H_{j-1/2}^{\left( 1\right) }\left( \varsigma \right) \right]  \notag \\
&=&-\tan \rho _{1}\left[ \sqrt{\left( \epsilon -1\right) /\left( \epsilon
+1\right) }\exp \left( -\psi _{1}\right) J_{j+1/2}\left( \varsigma \right)
H_{j+1/2}^{\left( 1\right) }\left( \varsigma \right) \right.  \notag \\
&&+\left. \sqrt{\left( \epsilon +1\right) /\left( \epsilon -1\right) }\exp
\left( \psi _{1}\right) J_{j-1/2}\left( \varsigma \right) H_{j-1/2}^{\left(
1\right) }\left( \varsigma \right) \right] ,  \label{massiveeq2} \\
\func{Re}E &>&0,\text{ }j=\pm 1/2,\pm 3/2,\ldots ,  \notag
\end{eqnarray}%
where $\varsigma =\gamma \sqrt{\epsilon ^{2}-1}.$ If $\psi
_{1}\longrightarrow \pm \infty $ (asymptotical approach to the bisectrix),
we have for the roots 
\begin{equation}
J_{j\mp 1/2}\left( \varsigma \right) =0,\text{ \ \ }\psi _{1}\longrightarrow
\pm \infty ,  \label{largepsiroots}
\end{equation}%
since the roots of the equation $H_{j\pm 1/2}^{\left( 1\right) }\left(
\varsigma \right) =0$ are extraneous ones.\ Solving Eq. (\ref{largepsiroots}%
), we obtain the Bessel function roots $\lambda _{n}^{\left( j\mp 1/2\right)
}.$ Therefore, 
\begin{equation}
\epsilon _{n}^{\left( j\mp 1/2\right) }=\sqrt{1+\left( \lambda _{n}^{\left(
j\mp 1/2\right) }/\gamma \right) ^{2}},\text{ \ \ }\psi _{1}\longrightarrow
\pm \infty  \label{deviation1}
\end{equation}%
This formula can be written in the form:%
\begin{equation}
E_{n}^{\left( j\mp 1/2\right) }=\sqrt{m^{2}+\left( \lambda _{n}^{\left( j\mp
1/2\right) }/r_{0}\right) ^{2}}  \label{deviation2}
\end{equation}

We see that appearing of non-zero mass shifts "arc's" bearing points
relative to the positions, determined by Eqs.. (\ref{rootminus}), (\ref%
{rootplus}), in the direction of the energy increase (see Fig. 14). This
shift is essential only for small $n$ since $\lambda _{n}^{\left( j\mp
1/2\right) }$ increases fast with $n$ increasing. When $m\neq 0,$ a
distribution of signs of Eq. (\ref{massiveeq2}) right-hand-side for various $%
m$ values at fixed $\rho _{1}$ is complicated. Arcs, in result, can be
formed both clock-wise and anti-clock depending on the $m$ value.

We have analyzed the resonances for the region I; the dependence of the
resonances distribution on the parameter $\gamma =r_{0}m$ in other regions
can be considered similarly.

We studied in the previous section a motion of resonances along the
figure-of-eight curves in the case of $m=0$, when the parameters $a,b$ are
varied along the circles Eq.(\ref{circle}) with various $\rho $ values. In
the non-zero gap case, we have found that a large enough mass destructs the
figure-of-eight structure for $\rho $ fixed; these figures associate into a
whole complicated aggregate (see Fig. 15).

\section{Conclusion}

In conclusion, we considered the bound and resonance electronic states for
the two-dimensional Dirac equation with the short-range perturbation. The
short-range perturbation is approximated by the delta function $\delta
\left( r-r_{0}\right) $ with different amplitudes in the upper and lower
electronic bands. We have in result local perturbations of the
potential-like and mass-like types.respectively with amplitudes $a$ and $b$.

We derived the characteristic equations for the bound and resonance states.
The characteristic equation is presented in forms convenient both for an
analysis of the bound and resonance states. Energy levels behavior in
dependence on the perturbation amplitudes was investigated both analytically
and numerically. A general picture of the electronic spectrum dependence on
the potential, and mass local perturbation is considered for various mass
and angular momentum values; the results are presented as diagrams in the $%
\left( a,b\right) $-plane. Absence of a threshold for forming of the bound
state in the vicinity of the upper bandgap edge is obtained for the lower
angular momenta $\left\vert j\right\vert =1/2$ which is in accord with the
general principles of quantum mechanics for the dimension 2+1.

A qualitative distinction of $a^{2}>b^{2}$ and $a^{2}>b^{2}$ cases is shown.
A countable set of eigenvalues existence domains in the $\left( a,b\right) $%
-plane is present in the former case, while monotonic approaching to the
asymptote takes place,$\ $when $b^{2}-a^{2}\longrightarrow \infty $. in the
latter one. Positions of asymptotes are determined by the parameter $\tanh
^{-1}\left( b/a\right) .$ Higher angular momenta are investigated as well.
Twisted eigenvalue domains with crossing boundaries are found. Domains of
eigenvalues are narrowing with the angular momentum increase and at the mass
value fixed. A zero-energy solution with vanishing imaginary part exists in
the zero-gap case.This solution has to be considered as a limit at $%
m\longrightarrow 0.$ \qquad \qquad

Behavior of resonances depends essentially on the sign of $\left(
a^{2}-b^{2}\right) .$ When $a^{2}>b^{2},$ resonance trajectories in the
energy complex plane have a form of closed curves circumscribed periodically
at moving from the coordinate origin to infinity in the ($a,b)$-plane.
Motion along the hyperbolae $\sqrt{a^{2}-b^{2}}=const$ maps onto the
countable set of arc-like trajectories in the energy complex plane. These
arcs bear on the real axis; positions of the bearing points are exactly
determined by the Bessel function roots. When $b^{2}-a^{2}>0,$ an increase
of $b^{2}-a^{2}$ forces the arcs to get flatter so that if $%
b^{2}-a^{2}\longrightarrow \infty ,$ the arcs asymptotically nestle up to
the real axis. The resonance width is exponentially small in this case. For
higher angular momenta $j,$\ the rate of this decrease is getting small due
to the factor $\left\vert j\right\vert ^{\left\vert j\right\vert }.$
Therefore, higher spherical harmonics do not contribute into the resonance
scattering.

Two kinds of the non-zero mass effect on the resonances behavior were found.
Firstly, positions of arc-bearing points shift. This effect is particularly
essential for lower roots. Secondly, eight-like figures resulted from
mapping of the circular motion in the $\left( a,b\right) $-plane onto the
energy plane transform into large aggregates.

The carried out analysis allowed us to describe both qualitatively and
quantitatively a distribution of the bound and resonance states in the
energy complex plane for any chosen distribution of the perturbation
amplitudes $a$ and $b$ and for various magnitudes of the mass and momentum.
The obtained results can be useful for understanding of the graphene
electronic properties.

\newpage
\section*{Figure captions}

%BeginExpansion
\begin{figure}
[h]
\begin{center}
\includegraphics[
height=3.5in, width=3.5in
]%
{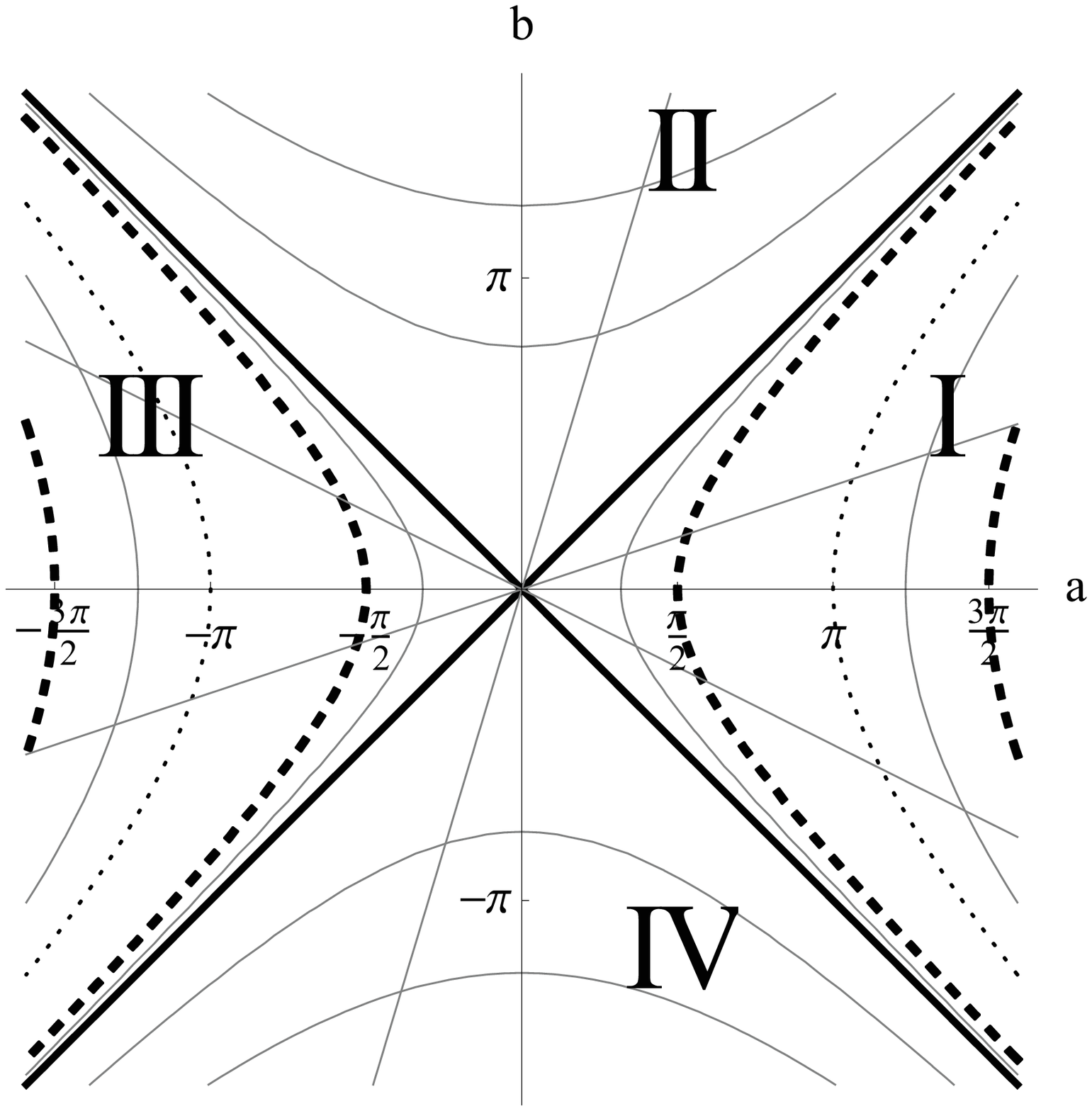}%
\caption{Regions of the perturbation amplitudes $\left( a,b\right) $-plane.
Bisectrices of the coordinate angles separating the $\left( a,b\right) $%
-plane are presented by the thick solid lines. The singular hyperbolae $\rho
_{1,3}=\pi /2+\pi n$ are presented by the thick dashed lines. Non-singular
hyperbolae are presented by the thin solid lines. The hyperbolae $\rho
_{1,3}=\pi n$ are presented by the thin dotted lines.}%
\end{center}
\end{figure}

\begin{figure}
[h]
\begin{center}
\includegraphics[
height=2in, width=2.33in
]%
{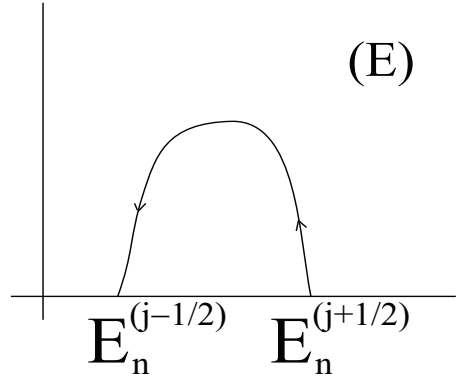}%
\caption{Schematic diagram of the arc-like resonance trajectory. $%
E_{n}^{\left( \lambda \pm 1/2\right) }$ are the roots of the Bessel
functions $J_{j\pm 1/2}\left( r_{0}E\right) .$}%
\end{center}
\end{figure}

\begin{figure}
[h]
\begin{center}
\includegraphics[
height=2in, width=2in
]%
{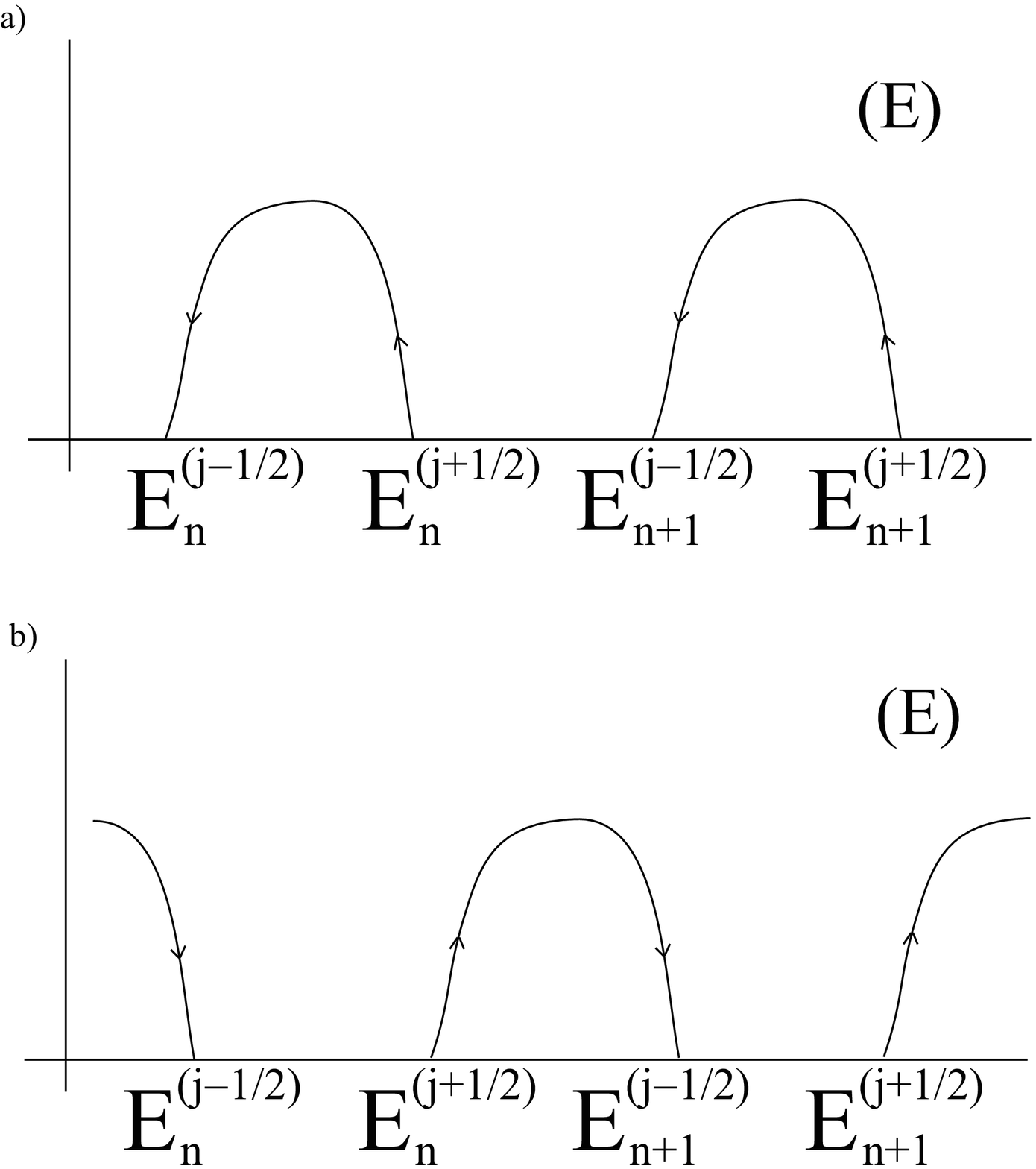}%
\caption{
Recommuting of the arc system in the region I, when $\rho _{1}$
crosses the singular hyperbolae $\rho _{1}=\pi /2+\pi n$, $n=0,1\ldots $
\\(a) $\pi n<\rho _{1}<\pi /2+\pi n.$
\\(b) $\pi /2+\pi n<\rho _{1}<\pi \left( n+1\right) .$
}%
\end{center}
\end{figure}

\begin{figure}
[h]
\begin{center}
\includegraphics[
height=2in, width=3in
]%
{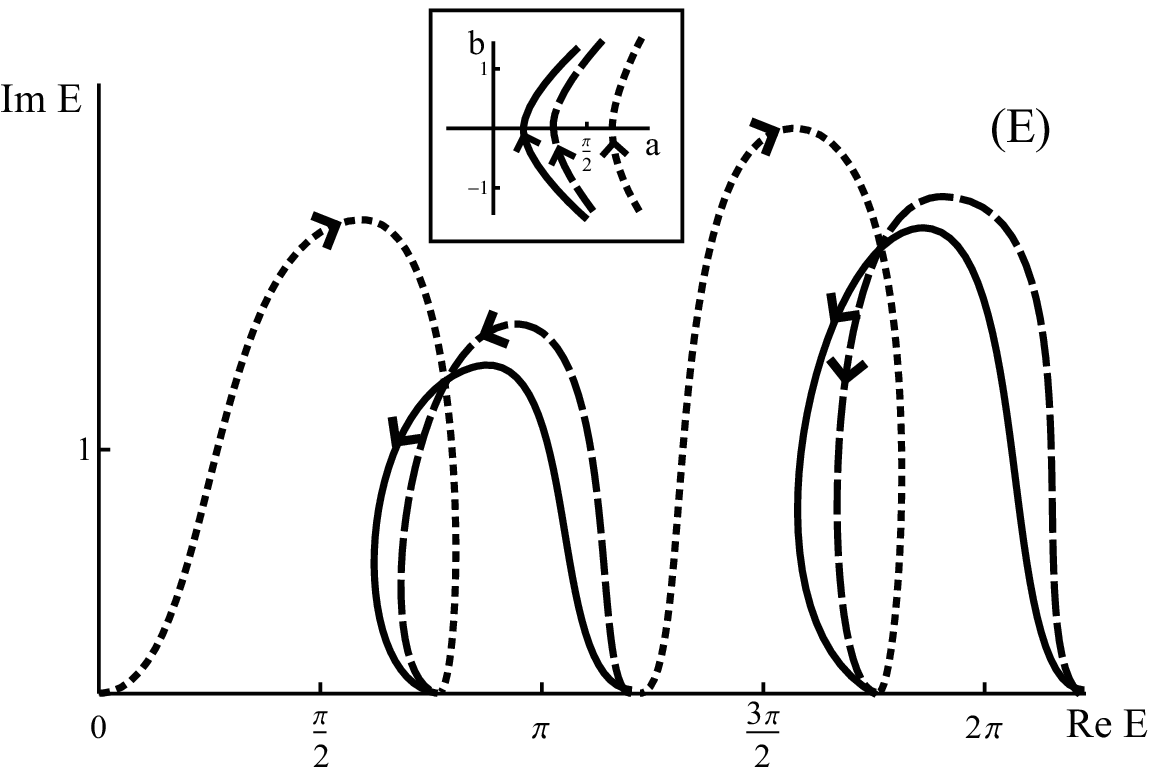}%
\caption{Numerically calculated resonance trajectories for various $\rho _{1}$
values in the region I. Solid lines: $\rho _{1}=0.8,$ dashed lines: $\rho
_{1}=1.4,$ dotted line $\rho _{1}=2.0.$ 
\\In the inset: Hyperbola paths in the $\left( a,b\right) $-plane. 
\\In the body of figure: Maps of the above paths onto the resonances
trajectories in the energy complex plane.}%
\end{center}
\end{figure}

\begin{figure}
[h]
\begin{center}
\includegraphics[
height=2in, width=3in
]%
{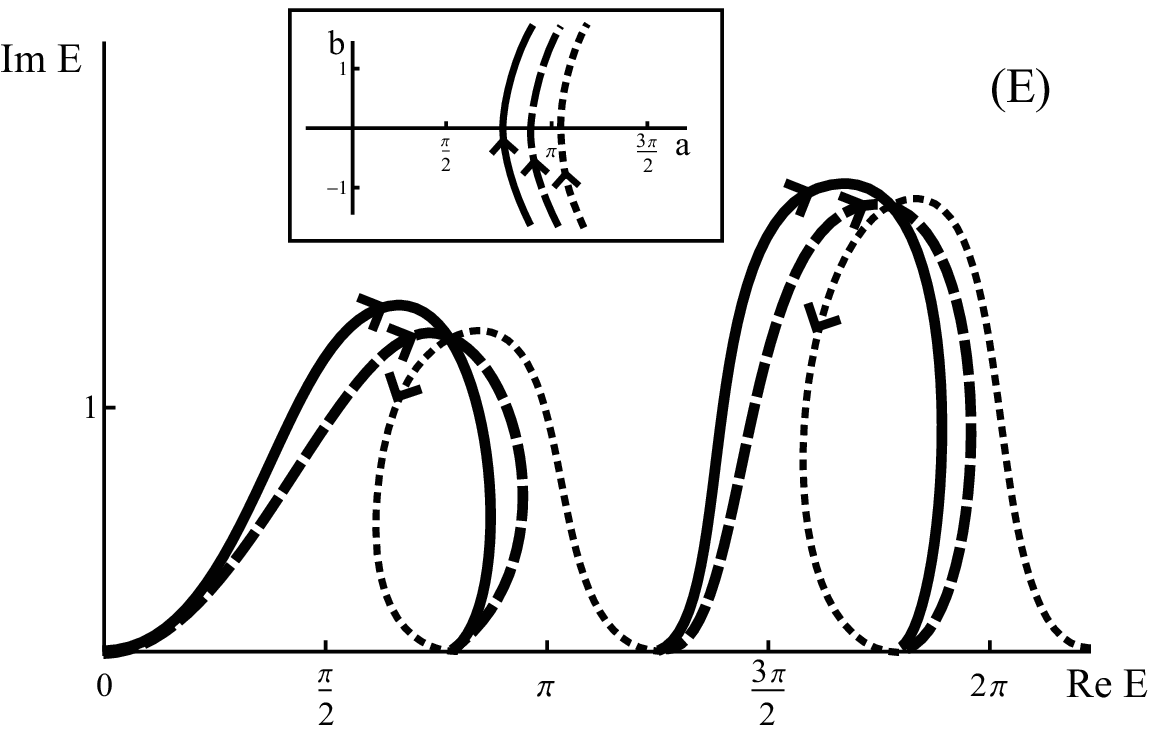}%
\caption{Numerically calculated resonance trajectories for various $\rho $
values in the region I. Solid lines: $\rho _{1}=2.5,$ dashed lines: $\rho
_{1}=3.0,$ dotted grey lines $\rho _{1}=3.5.$
\\In the inset: Hyperbola paths in the $\left( a,b\right) $-plane.
\\In the body of figure: Maps of the above paths onto the resonances
trajectories in the energy complex plane.}%
\end{center}
\end{figure}

\begin{figure}
[h]
\begin{center}
\includegraphics[
height=2in, width=3in
]%
{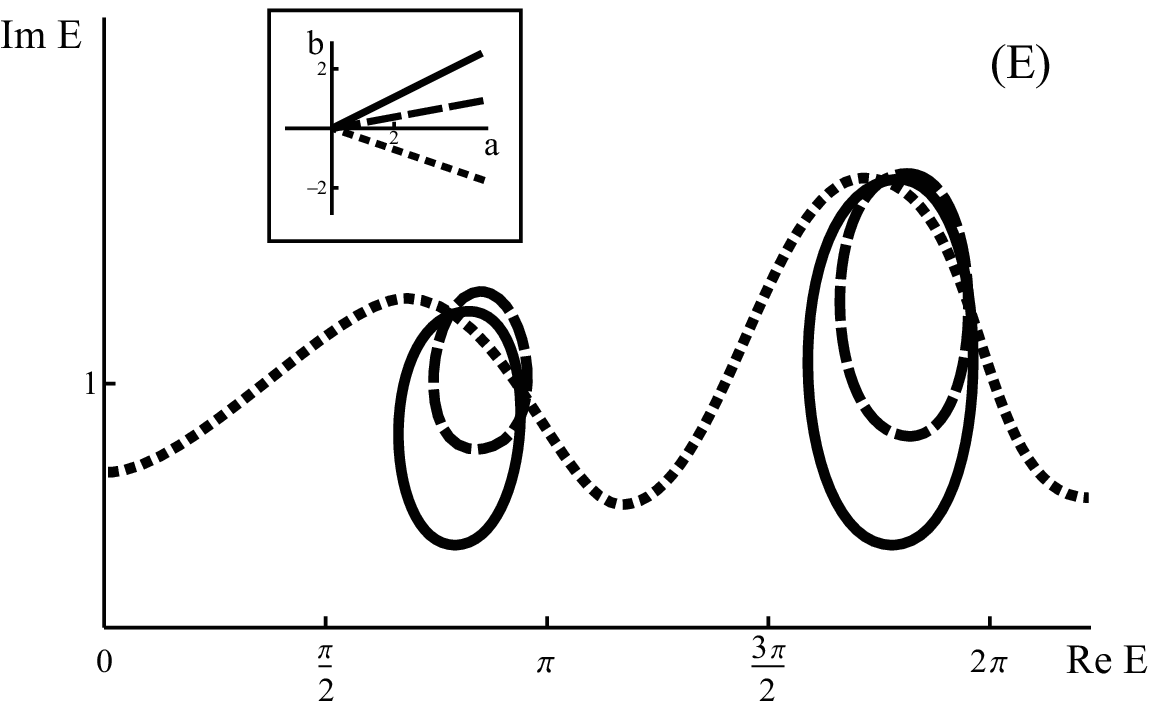}%
\caption{
Numerically calculated resonance trajectories for motion along the
rays in the $(a,b)$-plane in the region I. Solid lines: $\phi =\pi /7,$
dashed lines: $\phi =\pi /16,$ dotted lines: $\phi =-\pi /10.$
\\In the inset: Rays in the $\left( a,b\right) $-plane.
\\In the body of figure: Resonance trajectories in the energy complex plane.}%
\end{center}
\end{figure}

\begin{figure}
[h]
\begin{center}
\includegraphics[
height=2in, width=3in
]%
{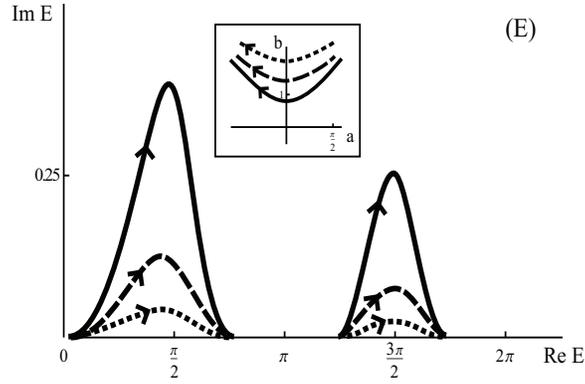}%
\caption{Numerically calculated resonance trajectories for various $\rho _{2}$
values in the region II. Solid lines: $\rho _{2}=0.8,$ dashed lines: $\rho
_{2}=1.4,$ dotted lines $\rho _{2}=2.0.$
\\In the inset: Hyperbola paths in the $\left( a,b\right) $-plane.
\\In the body of figure: Maps of the above paths onto the resonances
trajectories in the energy complex plane.}%
\end{center}
\end{figure}

\begin{figure}
[h]
\begin{center}
\includegraphics[
height=2in, width=3in
]%
{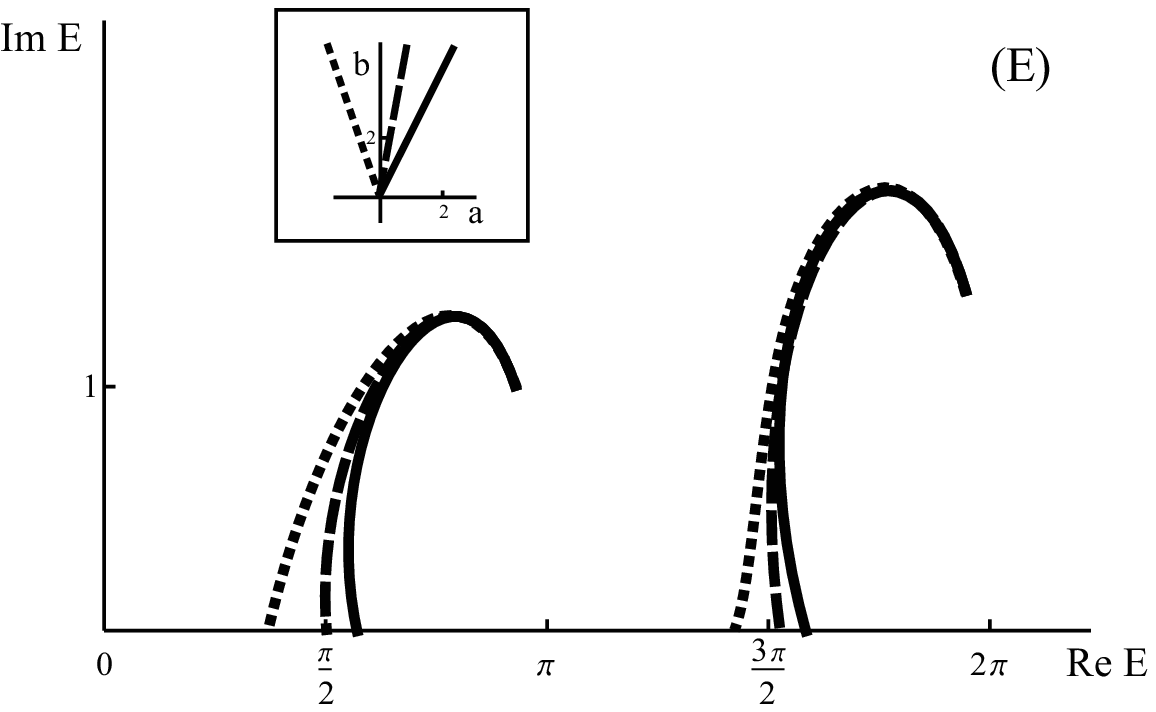}%
\caption{Numerically calculated resonance trajectories for motion along the
rays in the $(a,b)$-plane in the region II. Solid lines: $\phi =5\pi /14,$
dashed lines: $\phi =7\pi /16,$ dotted lines $\phi =3\pi /5.$
\\In the inset: Rays in the $\left( a,b\right) $-plane.
\\In the body of figure: Corresponding resonance trajectories in the energy
complex plane.}%
\end{center}
\end{figure}

\begin{figure}
[h]
\begin{center}
\includegraphics[
height=2in, width=3in
]%
{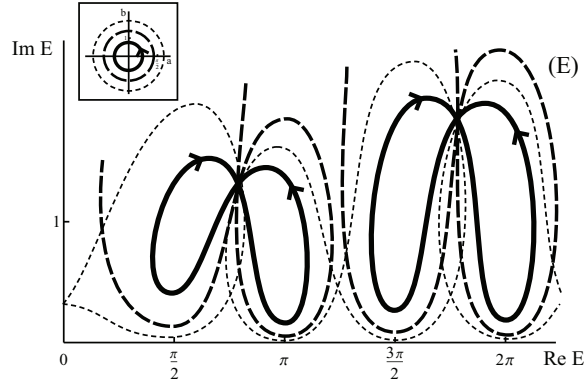}%
\caption{Numerically calculated resonance trajectories for motion along the
circles in the $(a,b)$-plane. Solid lines: $\rho =0.8,$ dashed lines $\rho
=1.4,$ dotted lines $\rho =2.0.$
\\In the inset: Circles in the $\left( a,b\right) $-plane.
\\In the body of figure: Corresponding figure-of-eight resonance trajectories
in the energy complex plane.}%
\end{center}
\end{figure}

\begin{figure}
[h]
\begin{center}
\includegraphics[
height=1.7in, width=3in
]%
{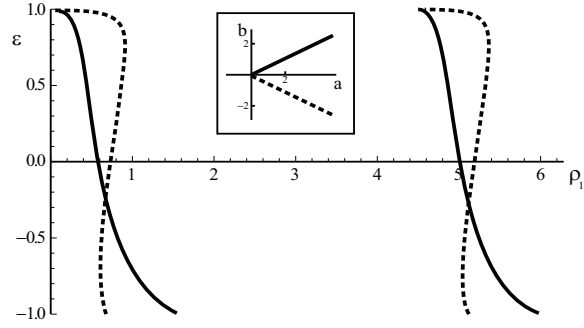}%
\caption{Eigenvalue dependence on the radial coordinate $\rho _{1}$ in the
region I for two angle magnitudes: solid line $\phi =\pi /6,$ dashed line $%
\phi =-\pi /6.$In the inset: Two rays in the $(a,b)$-plane with the above
angle values.}%
\end{center}
\end{figure}
\begin{figure}
[h]
\begin{center}
\includegraphics[
height=1.7in, width=3in
]%
{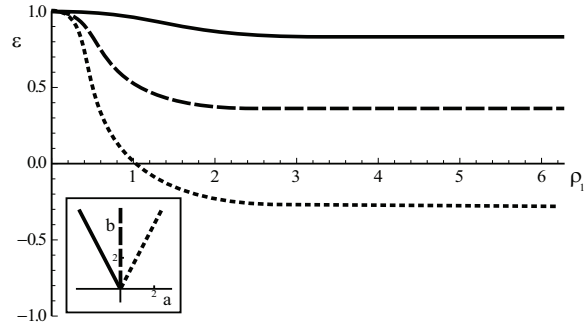}%
\caption{Eigenvalue dependence on the radial coordinate $\rho _{2}$ in the
region II for three angle values (corresponding rays are shown in the
inset): solid lines $\phi =2\pi /3,$ dashed lines $\phi =\pi /2,$ dotted
lines $\phi =-2\pi /3.$}%
\end{center}
\end{figure}
\begin{figure}
[h]
\begin{center}
\includegraphics[
height=3in, width=3in
]%
{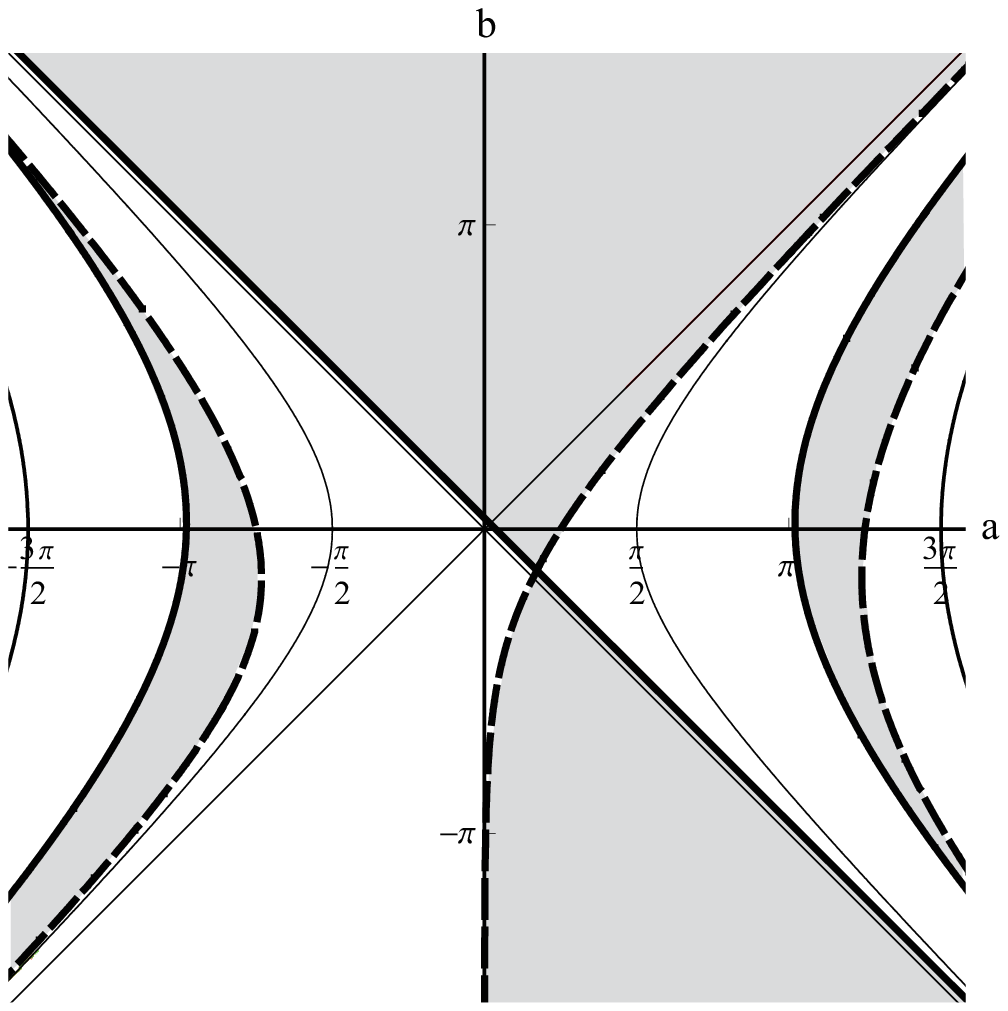}%
\caption{Diagram of bound states for $j=1/2$ and $\gamma =2.$ "Fronts" $%
\epsilon =+1$ are shown by solid lines, "fronts" $\epsilon =-1$ are shown by
dashed lines. Thin solid line hyperbolae are asymptotes of the "fronts".}%
\end{center}
\end{figure}
\begin{figure}
[h]
\begin{center}
\includegraphics[
height=3in, width=3in
]%
{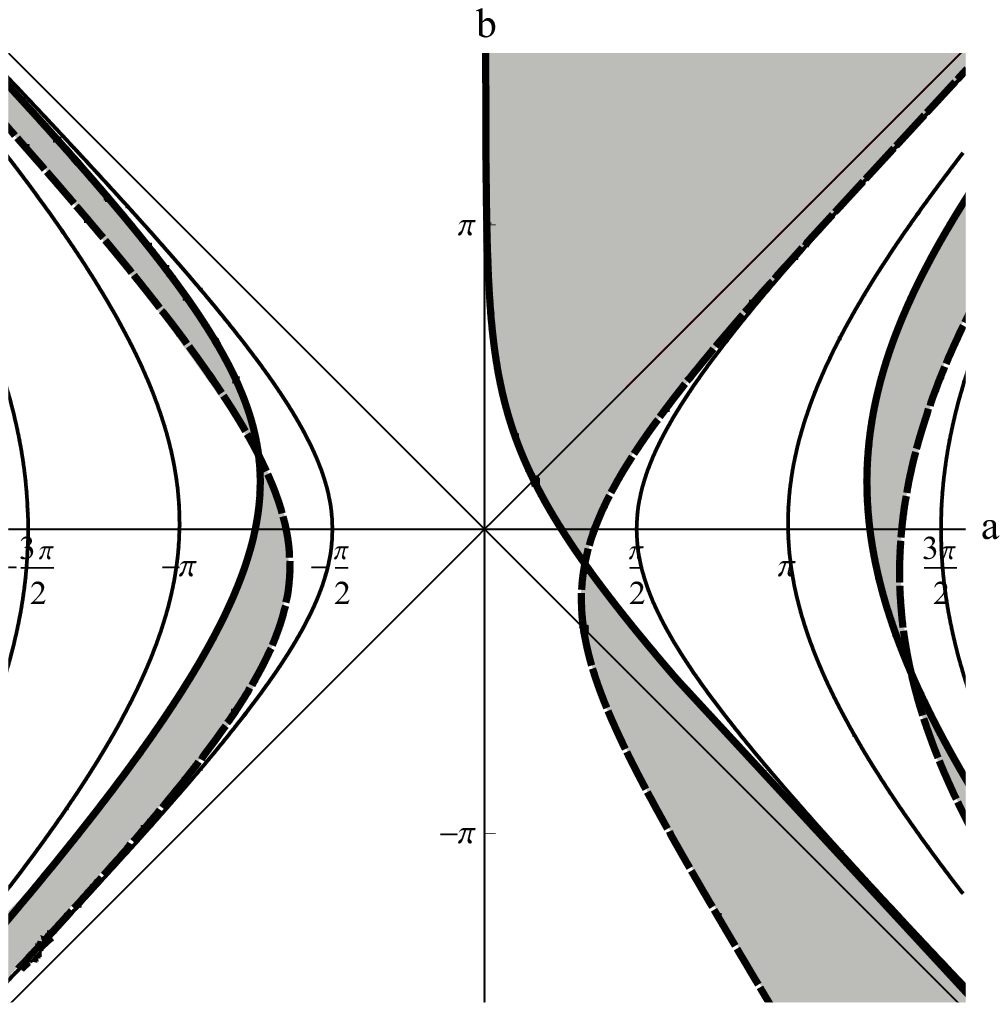}%
\caption{Diagram of bound states for $j=3/2$ and $\gamma =2.$ "Fronts" $%
\epsilon =+1$ are shown by solid lines, "fronts" $\epsilon =-1$ are shown by
dashed lines. Thin solid line hyperbolae are asymptotes of the "fronts".$.$}%
\end{center}
\end{figure}
\begin{figure}
[h]
\begin{center}
\includegraphics[
height=2in, width=3in
]%
{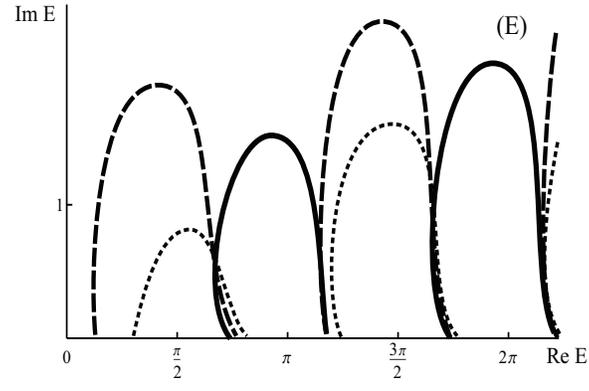}%
\caption{Arc-like resonance trajectory dependence on the mass (gap) value.
Solid lines: $m=0,$ dashed lines: $m=0.5,$ dotted line $m=1.0.$}%
\end{center}
\end{figure}
\begin{figure}
[h]
\begin{center}
\includegraphics[
height=2in, width=3in
]%
{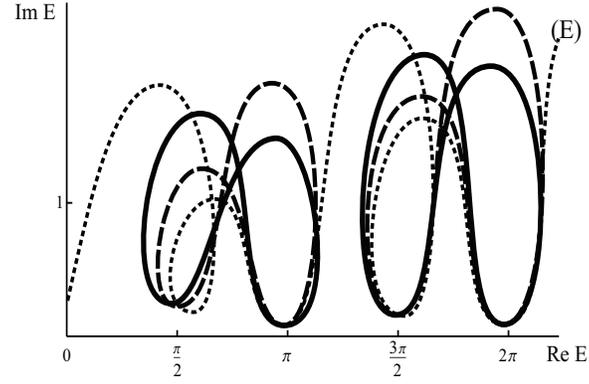}%
\caption{Figure-of-eight resonance trajectory dependence on the mass (gap)
value. Solid lines: $m=0,$ dashed lines: $m=0.25,$ dotted line $m=0.5.$}%
\end{center}
\end{figure}

%EndExpansion

\end{document}